%% file: FCCM.tex
\documentclass[conference]{IEEEtran}
\ifCLASSINFOpdf
\else
\fi

\usepackage{comment}
\usepackage{graphicx}
\usepackage{xcolor}
\usepackage{multirow}
\usepackage{amsmath,amssymb,amsfonts}
\usepackage{siunitx}
\usepackage{todonotes}
\usepackage{cite}
\usepackage{amsmath,amssymb,amsfonts}
\usepackage{algorithmic}
\usepackage{subfig}
\usepackage{makecell}
\usepackage{svg}
\usepackage{tablefootnote}
\usepackage[flushleft]{threeparttable}
\usepackage[symbol]{footmisc}
\usepackage{fixltx2e}
\usepackage{caption} 
\usepackage{soul}

\usepackage{url}

\usepackage{breakurl}
\usepackage[breaklinks]{hyperref}

\newcommand{\arraylength}[1][]{\textit{TB\textsubscript{len}}#1}
\newcommand{\Mexpansion}[1][]{\textit{M\textsubscript{p}}#1}
\newcommand{\Nexpansion}[1][]{\textit{N\textsubscript{p}}#1}
\newcommand{\Kexpansion}[1][]{\textit{K\textsubscript{p}}#1}

\begin{document}
%

\title{Efficient Approaches for GEMM Acceleration on Leading AI-Optimized FPGAs}

\author{\IEEEauthorblockN{Endri Taka\IEEEauthorrefmark{2}\textsuperscript{\textsection}, Dimitrios Gourounas\IEEEauthorrefmark{2}\textsuperscript{\textsection}, Andreas Gerstlauer\IEEEauthorrefmark{2}, Diana Marculescu\IEEEauthorrefmark{2}, Aman Arora\IEEEauthorrefmark{3}}
\IEEEauthorblockA{\IEEEauthorrefmark{2}\textit{The University of Texas at Austin, USA}, \IEEEauthorrefmark{3}\textit{Arizona State University, USA}}
\IEEEauthorblockA{\{endri.taka, dimitrisgrn, gerstl, dianam\}@utexas.edu, aman.kbm@asu.edu
}
}
\maketitle
\begingroup\renewcommand\thefootnote{\textsection}
\footnotetext{Authors contributed equally to this work.}
\endgroup

\begin{abstract}
FPGAs are a promising platform for accelerating Deep Learning (DL) applications, due to their high performance, low power consumption, and reconfigurability. 
Recently, the leading FPGA vendors have enhanced their architectures to more efficiently support the computational demands of DL workloads.
However, the two most prominent AI-optimized FPGAs, i.e., AMD/Xilinx Versal ACAP and Intel Stratix 10 NX, employ significantly different architectural approaches. 
This paper presents novel systematic frameworks 
to optimize the performance of General Matrix Multiplication (GEMM), a fundamental operation in DL workloads, by exploiting the unique and distinct architectural characteristics of each FPGA.
Our evaluation on GEMM workloads for int8 precision shows up to 77 and 68 TOPs (int8) throughput, with up to 0.94 and 1.35 TOPs/W energy efficiency for Versal VC1902 and Stratix 10 NX, respectively.
This work provides insights and guidelines for optimizing GEMM-based applications on both platforms, while also delving into their programmability trade-offs and associated challenges.




\end{abstract}


%
\IEEEpeerreviewmaketitle

\section{Introduction}
\label{sec:Introduction}

\input{Introduction/Introduction}

\section{Related Work}
\label{sec:Related_Work}
\input{Related_Work/Related_Work}

\section{FPGA Architectures Overview}
\label{sec:Hardware_Overview}

\input{Hardware_Overview/HW_Overview}

\section{GEMM Design \& Optimization}
\label{sec:GEMM_Design}
\input{GEMM_Design/GEMM_Design}
\section{Evaluation}
\label{sec:Results}

\input{Results/Results}

\section{Conclusion}
\label{sec:Summary_Future_Work}
\input{Summary-Future_work/Summary_Future_Work}
\section*{Acknowledgment}

We thank the anonymous reviewers for their insightful feedback, which improved the quality of the paper.
This work was supported in part by the 
National Science Foundation
CCF Grant No. 2107085, ExxonMobil Technology
and Engineering Company, agreement no. EM10480.36, iMAGiNE -
the Intelligent Machine Engineering Consortium at UT Austin, and a UT
Cockrell School of Engineering Doctoral Fellowship.



%

\newpage

\bibliographystyle{IEEEtran}
\bibliography{Bibliography.bib}

\end{document}

%% file: Introduction/Introduction.tex
The explosion of computational demands in Deep Learning (DL) workloads \cite{TPUv1_2017, LLM_explosion}, has resulted in the emergence of AI-optimized hardware solutions, including GPUs \cite{NVIDIA_A100, NVIDIA_H100, AMD_CDNA_3} and ASICs \cite{TPUv42021, MTIA_Meta_2023, Intel_Goya_inference, Amazon_Inferentia2}.
In addition, several AI-optimized FPGA solutions have also been proposed, both in industry \cite{langhammer2021stratix, Versal_AI_Engines_FPGA_paper, TensorTile,tensor_ram,Achronix_Speedster} and academia \cite{aman_tensor_slices_FPGA, aman_tensor_slices_TRETS,comefa_trets,math_no_hard,logic_block_mohd,embrace_div_dsp}. 
The two major FPGA vendors have adopted different directions in optimizing their FPGAs for DL. 
AMD/Xilinx introduced the 
Versal Adaptive Compute Acceleration Platform (ACAP) \cite{VC1902_hot_chips, Alok2020apocalypse}, comprising the novel AI Engine (AIE), along with reconfigurable logic (FPGA) and scalar processors (CPUs).
In contrast, Intel released the Stratix 10 NX~\cite{langhammer2021stratix}, maintaining the existing FPGA architecture, but replacing legacy DSP blocks with new AI Tensor Blocks (TBs).
The AIE is an out-of-fabric solution consisting of programmable vector processors that operate at high frequency.
In contrast, TBs are in-fabric blocks comprising multiple dot-product engines, operating at lower FPGA fabric frequencies.

The two hardware platforms employ substantially different architectural attributes to incorporate DL support.
In this work, we present systematic methodologies and novel optimization techniques to map General Matrix-Matrix Multiplication (GEMM) workloads on the aforementioned AI-optimized FPGAs,
highlighting the distinct architecture-specific design approaches required for each device.
We propose frameworks that aim to maximize the throughput and energy efficiency of GEMM on Versal and Stratix FPGAs, leading to maximal resource utilization.
This study focuses on GEMM, since it constitutes the core operation in many DL workloads, occupying up to 90\% of the total execution time \cite{adolf2016fathom, wang_gemm_breakdown}.

Table \ref{table_hardware} presents the characteristics of the two exemplar devices, where we showcase our proposed methods; the Versal VC1902 and Stratix 10 NX 2100.
Both are large chips with roughly equal number of logic elements/cells, and similar on-chip memory capacity. 
Additionally, both devices have nearly equal \textit{theoretical peak} throughput ($int8$) capabilities, under similar power envelopes (135 \emph{vs.} 143 TOPs, and 165 \emph{vs.} 125 W, for Versal~\cite{VCK_50000_installation} and Stratix~\cite{langhammer2021stratix}, respectively).

Besides their distinct architectures, the two devices also present differences in their DRAM technology and the manufacturing nodes (Table \ref{table_hardware}).
Versal has 5$\times$ lower bandwidth (BW) than Stratix. 
Moreover, Versal is manufactured in a 7\textit{nm} TSMC process, while Stratix uses 14\textit{nm} Intel.
The main focus of this paper is to provide a comprehensive evaluation of various aspects in GEMM optimization,
emphasizing architecture-specific methodologies, which are largely agnostic to DRAM and manufacturing technology.
Thus, we perform experiments on designs that operate within on-chip memory.
However, to enable a complete and thorough analysis we extensively examine off-chip memory considerations and requirements for both devices.
Our main contributions are summarized below:

\begin{table}[t]
\vspace{+2mm}
\centering
\resizebox{1.00\linewidth}{!}{
\begin{threeparttable}
 \centering
\setlength\tabcolsep{5.0pt}
\renewcommand{\arraystretch}{1.05}
\caption{Hardware platform characteristics.}
\begin{tabular}{c|c c c}
\Xhline{2.5\arrayrulewidth}
\multicolumn{2}{c}{\textbf{Device}} & \textbf{Versal VC1902} & \textbf{Stratix 10 NX 2100}\\
\hline
\hline
\multirow{4}{*}{\rotatebox[]{90}{\textbf{FPGA}}} 
& \textbf{Logic Elements/Cells}  & 1968K & 2073K \\
& \textbf{On-chip Memory}  & 20.5 MB & 16.75 MB\\
& \textbf{DSP Slices}  & 1968 & --\\
& \textbf{Tensor Blocks}  & -- & 3960\\
\hline
\multicolumn{2}{c}{\textbf{AIE Cores}} & 400 & -- \\
\multicolumn{2}{c}{\textbf{AIE Memory}} & 12.5 MB & -- \\
\hline
\multicolumn{2}{c}{\textbf{Processing System}} & ARM A72 + R5F & -- \\
\hline
\multicolumn{2}{c}{\textbf{DRAM Technology}} & DDR4 & HBM2\\
\multicolumn{2}{c}{\textbf{Peak DRAM BW}} & 102.4 GB/s & 512 GB/s \\
\hline
\multicolumn{2}{c}{\textbf{Theor. Peak TOPs (int8)*}} & 135 TOPs & 143 TOPs\\
\multicolumn{2}{c}{\textbf{Peak Power}} & 165 W & 125 W \\
\multicolumn{2}{c}{\textbf{Process}} & 7nm TSMC & 14nm Intel \\

\Xhline{2.5\arrayrulewidth}
\end{tabular}

\begin{tablenotes}
      \footnotesize
      \item* 
      Versal's throughput is primarily attributed to the AIE, providing a peak of 128 TOPs at 1.25 GHz.
      DSPs present only 7 TOPs at 600 MHz, hence are not considered in this work.
      Stratix 10 NX peak throughput is reported in~\cite{IntelStratix10NX_WhitePaper}, for operation at 600MHz.
\end{tablenotes}


\vspace{-0.55cm}
\label{table_hardware}
\end{threeparttable}
}
\end{table}

\begin{itemize}
\item For Versal, we leverage the state-of-the-art MaxEVA framework \cite{Taka2023MaxEVA}, and extend it
to incorporate an additional memory hierarchy level utilizing the Versal FPGA's on-chip resources. 
We maximize performance via design space exploration (DSE) and analytical modeling, and we propose a novel RAM optimization scheme to overcome severe limitations of Vitis High-Level Synthesis (HLS).

\item For Stratix, we develop a novel framework to design, map and optimize a configurable GEMM accelerator by exploiting the device's in-fabric TBs.
Our framework involves extensive DSE and analytical modeling to maximize GEMM performance.

\item Demonstration of our frameworks on various GEMM workloads for $int8$ precision, showing throughput up to 77 and 68 TOPs with 100\% AIE and 91\% TB utilization for Versal and Stratix, respectively. We achieve up to 0.94 and 1.35 TOPs/W energy efficiency, with 88\% and 94\% on-chip memory for Versal and Stratix, respectively.

\item We provide notable insights and guidelines for GEMM optimization, programmability aspects, architectural attributes, and limitations on both AI-optimized FPGAs.
    
\end{itemize}

%% file: Related_Work/Related_Work.tex
Several prior works leverage the Versal ACAP architecture across multiple application domains.
In particular, CHARM \cite{charm2023fpga, charm_DAC23} automates the process of GEMM acceleration on Versal ACAP.
Their experimental results on the VC1902 exhibit higher energy efficiency of up to 7.2$\times$ and 1.7$\times$ compared to traditional FPGAs (AMD/Xilinx U250) and GPUs (NVIDIA A100), respectively.
MaxEVA~\cite{Taka2023MaxEVA} is another framework that accelerates GEMM on Versal AIE, while achieving up to 2.19$\times$ higher throughput and 20.4\%  higher energy efficiency compared to CHARM.
Additional research on Versal focuses on accelerating specific DL workloads, such as Convolutional Neural Networks (CNNs) \cite{XVDPU_AIE_FPL22}, \cite{CNN_acc_Versal_FPL22}, and Graph Neural Networks (GNNs) \cite{H_GCN_FPL2022, chen2023exploiting}.
Other works include AIE compilers \cite{AIE_compiler_HPEC20}, arbitrary precision integer multiplication \cite{yang2023aim}, as well as acceleration of atmospheric simulations \cite{sparta_ICS23, Stensil_AIE_FPGA23}.
Considering all prior work, MaxEVA is the state-of-the-art GEMM implementation, although only targeting small matrix sizes that fit within the Versal AIE.
In this work, we extend MaxEVA to support arbitrary GEMM sizes, by implementing an additional level of memory hierarchy on Versal's FPGA.

The NX architecture was introduced in~\cite{langhammer2021stratix}, including a discussion of the TB operating modes and design trade-offs. This work also presents a TB design used for General Matrix-Vector (GEMV) and GEMM operations.
Multiple other works have targeted the Stratix 10 NX for DL.
In \cite{boutros2020beyond}, a GEMV accelerator is mapped on AI TBs and incorporated into an enhanced Brainwave NPU overlay~\cite{brainwave_NPU}. 
They show a speedup of up to 3.5$\times$ compared to all prior works on FPGA-based acceleration of Recurrent Neural Networks (RNN)~\cite{Guan_LSTM_FPGA,Vladimir_FINNL,Que_RNN_FPGA}, as well as the baseline NPU with legacy DSPs~\cite{Stratix10GX_NPU}. 
Additionally, in~\cite{hpipex_NX}, NX was utilized to enhance the CNN HPIPE accelerator~\cite{HPIPE_baseline}.
They demonstrate a 4$\times$ speedup over prior FPGA accelerators for CNN.
In~\cite{denseint16_NX}, a method to assemble higher than $int8$ precision multipliers is presented on the NX. 
Finally, in~\cite{WaveNet_NX}, 
a speech-generation model is implemented on Stratix 10 NX, substantially outperforming a V100 GPU implementation.
In this work, we develop a framework, which includes a detailed, systematic approach for automatically generating a configurable GEMM accelerator on Stratix 10 NX.
Furthermore, we perform an extensive DSE and we explore various trade-offs in GEMM design, which are not thoroughly examined in prior work targeting the NX device.






%% file: Hardware_Overview/HW_Overview.tex

\begin{figure}[tbp]
\vspace{-0.05cm}
\centering
\includegraphics[width=0.86\linewidth]{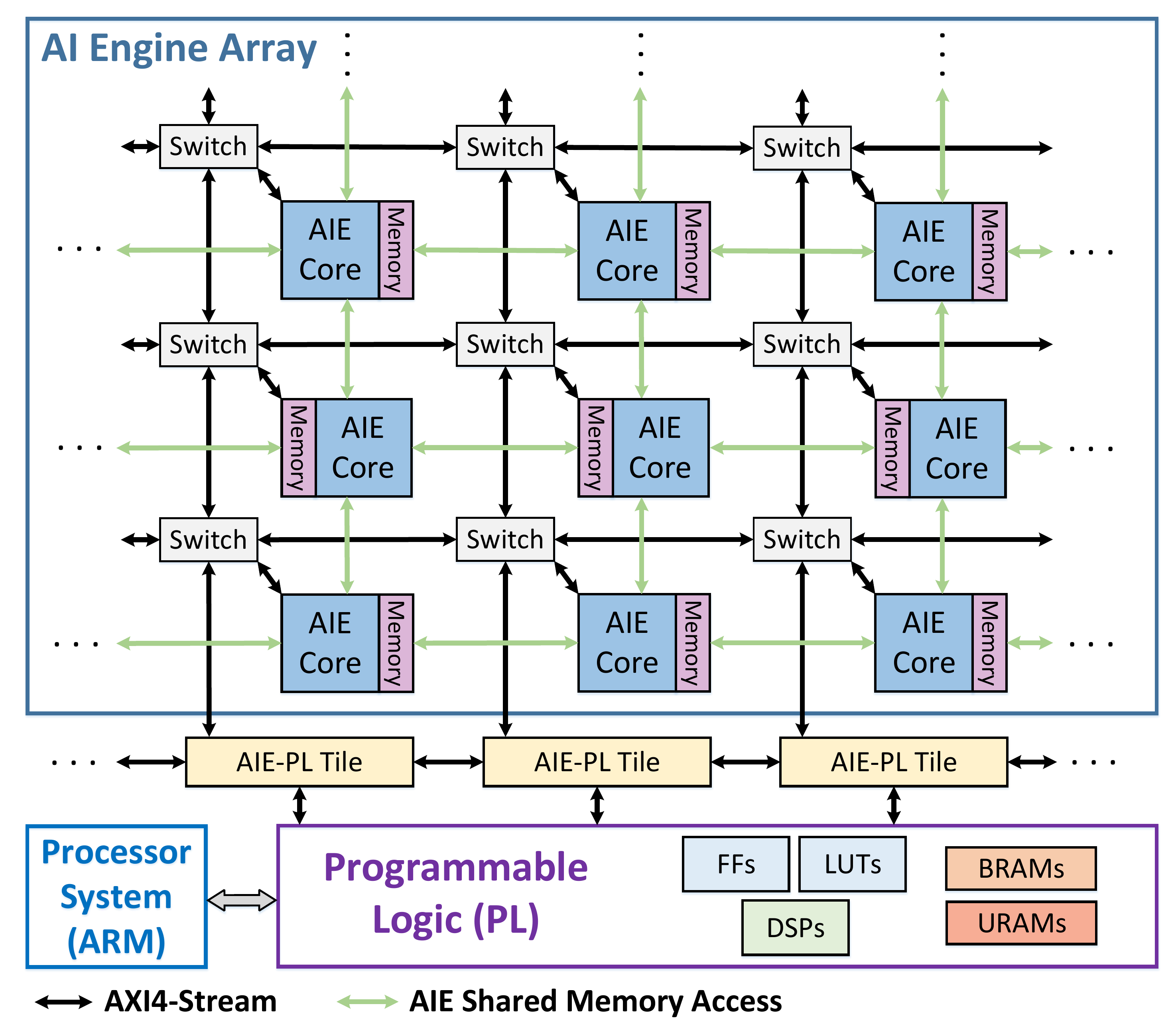}
\caption{Versal ACAP architecture.}
\label{fig:Versal_ACAP_Architecture}
\vspace{-0.50cm}
\end{figure}

\subsection{Versal ACAP Architecture}
\label{subsec:Versal_architecture}


The architecture of the Versal ACAP is depicted in Fig. \ref{fig:Versal_ACAP_Architecture}.
The Versal ACAP comprises the Processor System (PS), the Programmable Logic (PL), as well as the novel AIE array \cite{Versal_ACAP_design_guide}.
The PS consists of scalar ARM processors, while
the PL includes the traditional FPGA resources, \emph{e.g.}, LUTs, FFs, DSP slices, and on-chip memory resources (BRAMs/URAMs).



The Versal AIE is a 2D array consisting of identical AIE tiles.
Each tile includes an AIE core, a memory unit and an interconnect (switch)~\cite{Versal_acap_architecture_manual}.
The AIE cores are architected as VLIW programmable processors featuring vector (SIMD) units.
The AIE array provides three levels of parallelism.
First, \emph{instruction-level} parallelism is realized by executing up to 7 instructions every clock cycle (7-way VLIW).
Second, \emph{data-level} parallelism is achieved through vector operations, where multiple data can be processed each clock cycle (SIMD).
Third, \emph{spatial-level} parallelism is attained via the concurrent execution of multiple AIE cores (up to 400).
Communication between different AIE cores is achieved by local memory sharing access for neighboring cores, or by programmable switches for distant cores (Fig. \ref{fig:Versal_ACAP_Architecture}).
The switches can be configured statically (at compile time) for circuit-switching, or dynamically for packet-switching.
Circuit-switching is more efficient, ensuring deterministic latency, as opposed to non-deterministic latency associated with packet-switching \cite{Versal_acap_architecture_manual}.

The AIE array communicates efficiently with the PL via the dedicated AIE-PL tiles, located on the last row of the AIE, as shown in Fig. \ref{fig:Versal_ACAP_Architecture}.
These tiles provide AXI4-Streaming interface with the PL, while also supporting clock domain crossing between the AIE and the PL.
The Versal ACAP additionally includes a Network-on-Chip (NoC) (not shown in Fig. \ref{fig:Versal_ACAP_Architecture}), to enable flexible communication throughout the entire chip.






An AIE kernel running on a single AIE core can be programmed in high-level C/C++~\cite{AIE_API_user_guide}, or low-level SIMD intrinsics \cite{AI_Engine_programming_guide}.
The mapping of multiple kernels on the AIE array is realized through the Adaptive Data Flow (ADF) graph modeling.
The nodes in the ADF correspond to AIE kernels and the edges represent connections between them \cite{AI_Engine_programming_environment}.
The PL can be programmed in C/C++ using Vitis 
HLS \cite{Vitis_HLS_guide} or low-level RTL.
Finally, AMD/Xilinx provides the Vitis V++ tool~\cite{Vitis_software_guide} to integrate the AIE graph system and the PL kernels.




\subsection{Stratix 10 NX Architecture}
\label{subsec:Stratix_10_Nx_architecture}
\begin{figure}[t]
\centering
\includegraphics[width=0.85\columnwidth]{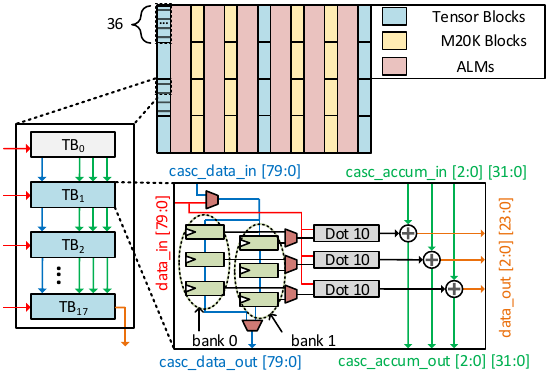}
\caption{Architecture of Stratix 10 NX Tensor Blocks.}
\label{fig_tensor_block_diagram}
\vspace{-0.50cm}
\end{figure}

The Intel Stratix 10 NX 2100 device's PL comprises ALMs, FFs, M20K blocks (Intel's BRAMs, 20Kbit in size~\cite{Intel_M20K_guide}) and the AI-optimized TBs. TBs replace the traditional variable-precision DSP blocks (Fig.~\ref{fig_tensor_block_diagram}), by dropping many of their legacy, high-precision operating modes and replacing them with scalar, vector, and tensor operating modes for several DL-optimized data types. 
The TB maintains the interface of the legacy DSP. 
In this work, we focus on the Tensor $int8$ mode, which supports three 10-element 8$\times$8 signed 
dot-product operations and three 32-bit 
additions for accumulation on partial products.
TBs in an FPGA column are \textit{physically} grouped in chains, each consisting of 36 TBs.
The TBs within a chain are cascaded using dedicated wires to propagate operands and accumulated products. Neighboring chains are not cascaded. Within a chain, TBs can be \textit{logically} grouped in independent \textit{arrays} of configurable length.

Fig.~\ref{fig_tensor_block_diagram} illustrates an example of an \textit{array} of 18 TBs that lies within a TB chain, as well as a simplified block diagram of the 
TB operating in the $int8$ mode. Each TB contains two banks of three 
ping-pong registers for storing operands ($bank$ $0, 1$), where each register holds ten 8-bit values.
Moreover, it includes three 10-element dot-product engines ($Dot$ $10$). 
The first input to each dot-product engine comes from its corresponding register, while the second input is broadcast from the 80-bit wide $data\_in$ port to all three engines. Finally, three independent 32-bit fixed-point adders are responsible for adding the generated dot-products to the cascade input from the previous TB. As shown in Fig.~\ref{fig_tensor_block_diagram}, accumulation results are propagated along the TB \textit{array} through cascade connections between a block's $casc\_accum\_out$ port and the following block's $casc\_accum\_in$ port. The latency of the dot-product calculation plus the cascade accumulation is equal to two cycles. The \textit{array's} final TB outputs are exposed through three 24-bit wide $data\_out$ ports.

There are three methods for loading operands in an \textit{array's} TB registers: (i) parallel load mode, (ii) side load mode and (iii) cascade mode. In this work, we focus on the cascade mode, since it leads to less routing congestion, as mentioned in~\cite{langhammer2021stratix}. We refer the reader to~\cite{langhammer2021stratix} for a detailed description of the other modes and their trade-off analysis. In cascade mode, $TB_0$ (Fig.~\ref{fig_tensor_block_diagram}) performs no computation and acts only as a loading port, where operands enter the array through its 80-bit $data\_in$ port. These operands are then propagated to subsequent TBs ($TB_1$--$TB_{17}$ in Fig.~\ref{fig_tensor_block_diagram}) from the $casc\_data\_out$ port of one TB to the next's $casc\_data\_in$ port and eventually stored in the TB registers. This requires three cycles per TB, leading to longer loading latencies as the array grows in size. However, computation can occur concurrently due to the ping-pong registers, allowing to hide the loading latency.

While the TBs support many modes of operation, there is no available support for 
programming them using Intel HLS~\cite{Intel_HLS_guide} or other high-level tools.
Hence, the developer is responsible for generating a TB-based design exclusively in RTL.

%% file: GEMM_Design/GEMM_Design.tex


\subsection{GEMM Implementation on Versal ACAP}
\label{subsec:GEMM_Architecture_Versal}


We leverage the MaxEVA open-source code \cite{maxeva_github}, and extend it to include on-chip buffers in the PL, tiling logic, as well as Load/Store units to communicate with DDR.
The PL is designed using Vitis HLS, as extensively used in prior works on Versal to enhance productivity
\cite{charm2023fpga, H_GCN_FPL2022, Stensil_AIE_FPGA23, CNN_acc_Versal_FPL22}.
Optimization of the PL design is attained through analytical modeling for maximization of on-chip data reuse (to reduce DDR BW requirements, as VC1902 has limited BW, Table \ref{table_hardware}).
Moreover, this method effectively resolves severe memory over-utilization issues caused by Vitis HLS.
In the following sections, we provide a brief overview of the MaxEVA AIE design, and we elaborate on the PL design and optimization.

\subsubsection{GEMM Multi-Level Tiling Scheme}
\label{subsubsec:GEMM_Tiling_Versal}

\begin{figure}[tbp]
\centering
\vspace{-0.4cm}
\includegraphics[width=0.80\linewidth]{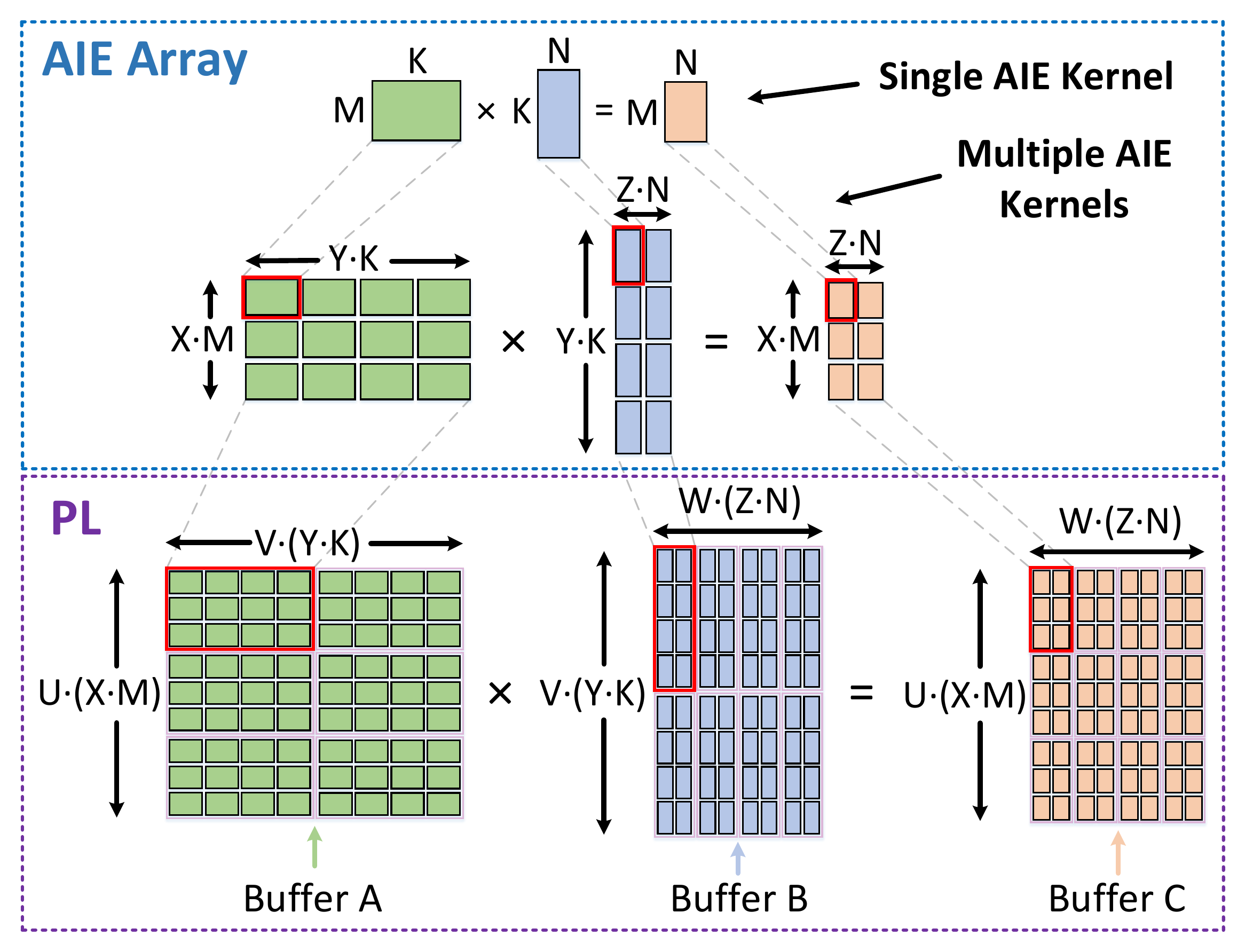}
\caption{Multi-level tiling scheme for GEMM on Versal ACAP.}
\label{fig:GEMM_tiling_Versal}
\vspace{-0.50cm}
\end{figure}

Fig. \ref{fig:GEMM_tiling_Versal} illustrates the tiling scheme used in MaxEVA for the AIE, as well as our tiling method on PL.
Each AIE core executes a Matrix Multiplication (MatMul) kernel of $M$$\times$$K$$\times$$N$ size (\textit{first tiling level}).
The parameters $X, Y, Z$ determine the multiple MatMul kernels running on the entire AIE array, as discussed in the next section (\textit{second tiling level}).
We incorporate an additional level of buffering in the PL, introducing three new parameters: $U, V, W$ (\textit{third tiling level}).
The AIE-specific parameters ($X, Y, Z, M, K, N$) are optimized by utilizing the MaxEVA framework, while the PL-specific parameters ($U, V, W$) are optimized using our proposed PL optimization procedure (introduced in Sec. \ref{subsubsec:PL_Optimization}).
All the aforementioned parameters determine the
$A$, $B$ and $C$ matrix sizes supported out of on-chip memory in the PL, as depicted in Fig. \ref{fig:GEMM_tiling_Versal}.
Therefore, we define two GEMM sizes.
First, the \textit{compute} GEMM size, \emph{i.e.}, $(X \cdot M) \times (Y \cdot K) \times (Z \cdot N)$, running on the AIE array.
Second, the \textit{native} buffer size, \emph{i.e.}, $(U \cdot X \cdot M) \times (V \cdot Y \cdot K) \times (W \cdot Z \cdot N)$, of the data stored in the on-chip buffers inside the PL.


\subsubsection{GEMM Mapping on AIE Array}
\label{subsubsec:GEMM_Mapping_AIE}

The limited number of AIE-PL tiles on Versal devices (notably VC1902 has only 39 AIE-PL tiles out of 50 columns in the AIE \cite{Versal_AI_DC_AC_switching}), is one of the main challenges in GEMM design.
To overcome the limited PL Input/Output (PLIO) bottleneck, MaxEVA utilizes the following two techniques \cite{Taka2023MaxEVA}.
First, the number of input PLIO ports is reduced by broadcasting input data to multiple AIEs.
Second, the output PLIO ports are decreased by performing adder tree reduction (via Add kernels) on the AIE.
This approach exploits only the most efficient circuit-switching AIE mechanism, as opposed to packet-switching used in \cite{charm2023fpga, charm_DAC23}.

In the upper part of Fig. \ref{fig:GEMM_design_AIE_PL} (AIE array), we present a high-level diagram of MatMul and Add kernels on the AIE.
Notice the groups of $Y$ MatMul kernels along with their corresponding adder trees ($Y-1$ Add kernels).
There exist $X \cdot Z$ such groups, all executing in  parallel.
Each MatMul kernel is mapped to a separate AIE core.
All Add kernels of a group (adder tree) are mapped to a single AIE core.
A total of $X \cdot Y\cdot Z$ AIE cores execute the MatMul kernels, and $X \cdot Z$ AIE cores execute the Add kernels.
Regarding the PLIOs, $X \cdot Y$ and $Y \cdot Z$ input ports are required for matrices $A$ and $B$, respectively, in addition to $X \cdot Z$ output ports for matrix $C$.

MaxEVA proposes two AIE kernel placement patterns, referred to as $P1$ and $P2$, to leverage the most efficient local data sharing mechanism of the AIE (Fig. \ref{fig:Versal_ACAP_Architecture}), and thus, avoid routing congestion.
$P1$ denotes a closely-located placement of each group of $Y=4$ MatMul kernels and their corresponding adder trees (Fig. \ref{fig:GEMM_design_AIE_PL}).
Similarly, $P2$ denotes a pattern for $Y=3$ (refer to \cite{Taka2023MaxEVA} for more details on these placement patterns).



\subsubsection{PL Implementation}
\label{subsubsec:PL_Implementation}
\begin{figure}[tbp]
\vspace{-0.40cm}
\centering
\includegraphics[width=0.945\linewidth]{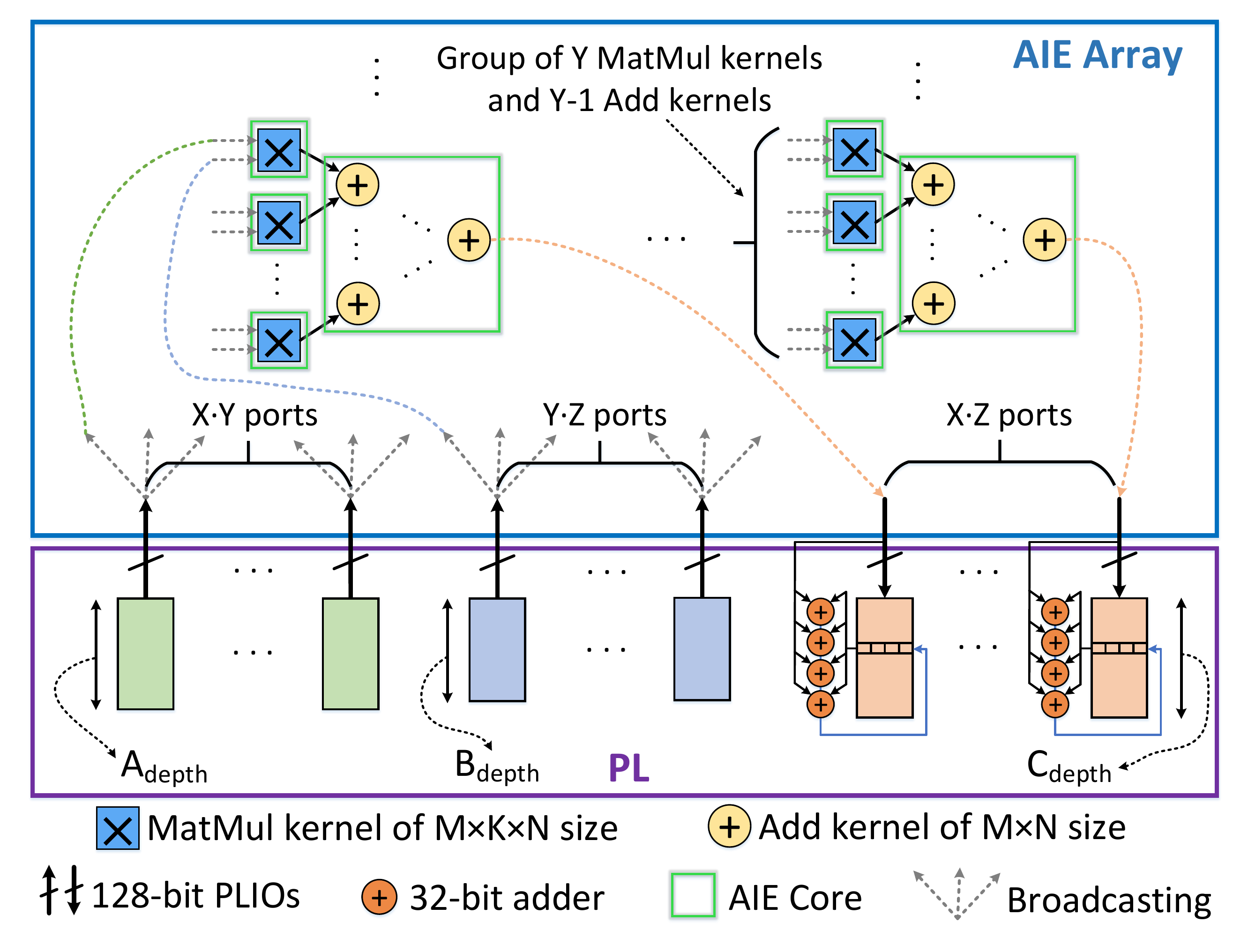}
\vspace{-0.2cm}
\caption{GEMM accelerator design on Versal AIE and PL.}
\label{fig:GEMM_design_AIE_PL}
\vspace{-0.50cm}
\end{figure}

The lower part of Fig. \ref{fig:GEMM_design_AIE_PL} (PL) shows a high-level block diagram of the PL design.
The 
$A$, $B$ and $C$ matrices are stored in on-chip buffers, exploiting the PL BRAM and URAM resources.
To provide sufficient bandwidth to/from the AIE, each buffer is partitioned ($HLS$ $pragma$ $array\_partition$), to exactly match the corresponding AIE PLIO ports.
Although not shown in Fig. \ref{fig:GEMM_design_AIE_PL} for simplicity, we employ  double-buffering to effectively overlap GEMM computation with external off-chip DDR communication.

We set the PLIO width to 128-bits to ensure rate matching between AIE and PL without performance loss \cite{Taka2023MaxEVA, AI_Engine_programming_guide}.
In addition, while the data type of the $A$ and $B$ buffers is $int8$, all accumulations are performed in 32-bits.
Thus, when sending data from the $A$ and $B$ buffers to the AIE, we concatenate 16 8-bit values for each input PLIO to form a 128-bit vector.
In contrast, when receiving data from the output PLIOs of the AIE, we pack 4 32-bit values and store them in the $C$ buffer.
The \emph{logical} size of the PL buffers is shown in Fig. \ref{fig:GEMM_tiling_Versal}.
However, in the \emph{physical} implementation, the buffers have a width of 128-bits to match the PLIO width, and are partitioned into smaller buffers
(Fig. \ref{fig:GEMM_design_AIE_PL}).
The partition factors ($\{A,B,C\}_{part}$) and depths ($\{A,B,C\}_{depth}$) of the buffers are expressed as:
\begin{align}
&A_{part} = 2 \cdot X \cdot Y, \quad A_{depth} = U \cdot V \cdot M \cdot K /16 \label{eq:A_buffer_part_depth}\\
&B_{part} = 2 \cdot Y \cdot Z, \quad B_{depth} = V \cdot W \cdot K \cdot N /16 \label{eq:B_buffer_part_depth}\\
&C_{part} = 2 \cdot X \cdot Z, \quad C_{depth} = U \cdot W \cdot M \cdot N /4 \label{eq:C_buffer_part_depth}
\end{align}
We multiply the partition factor by 2 for double-buffering. We also divide the depth by 16 and 4 to match the 128-bit packing.



Partial results from the AIE are accumulated in the PL to handle the reduction across tiles in the 
$V \cdot Y \cdot K$ dimension in Fig. \ref{fig:GEMM_tiling_Versal}, or for larger matrices if needed.
As depicted in Fig. \ref{fig:GEMM_design_AIE_PL}, we implement  4 32-bit adders in soft logic for each partitioned $C$ buffer ($4 \cdot X \cdot Z$ adders in total, all executing in parallel).
Each new partial result from every PLIO port (AXI4-Stream interface) is accumulated to its corresponding $C$ buffer address every clock cycle.
This PL logic is 
pipelined ($pragma$ $pipeline$) 
with an Initiation Interval (II) 
of 1, such that in every clock cycle a new partial result can be accumulated.
To this end, one load (for the current PLIO values) and one store operation (for the previous PLIO values) is required for every partitioned $C$ buffer.
Thus, we configure the $C$ buffers in \emph{simple} dual-port mode \cite{Versal_acap_memory_resources_manual, Vitis_HLS_guide} ($pragma$ $bind\_storage$).
This ensures a stall-free PL implementation that does not introduce any throughput degradation in the entire design.

In contrast, the input buffers $A$, $B$ are configured in single-port mode, since either only a load (send data to AIE) or store operation (receive data from DDR) is required in each cycle for double-buffering.
Moreover, Load/Store units are implemented in the PL (not shown in Fig. \ref{fig:GEMM_design_AIE_PL}) to communicate with DDR.
Finally, it is important to note that our implementation is symmetric in terms of the first and last dimensions in both the \textit{compute} GEMM size and the \textit{native} buffer size (see Sec. \ref{subsubsec:GEMM_Tiling_Versal} for definitions and Sec. \ref{subsubsec:GEMM_performance_versal} for evaluation).

\subsubsection{Memory Optimization Strategy}
\label{subsubsec:PL_Optimization}

To maximize the data reuse of the on-chip buffers, we propose an optimization methodology based on analytical modeling.
Our model utilizes the multiple configurations of BRAMs and URAMs to identify the optimal values of the $U, V, W$ parameters, as well as the buffer mapping to BRAMs and URAMs.
Although Vitis HLS includes the capability to automatically map buffers to memory resources ($impl$=$AUTO$ in $pragma$ $bind\_storage$), we found that it fails to find an operational mapping in several cases (Sec. \ref{subsubsec:PL_model_estimation}).
This automatic mapping generated by Vitis HLS leads to severe over-utilization of memory resources, and thus, failure to Place and Route (PnR). 
To the best of our knowledge, this HLS limitation has not been identified in any prior work. 
Several works have focused on optimizing the logical-to-physical memory mapping during the synthesis/PnR phases \cite{memory_mapping_slr, memory_mapping_old, memory_mapping_power}, while others \cite{memory_mapping_chow} propose tools to assist designers at the user-level 
with automatic memory mapping targeting BRAMs, but not URAMs.
Our approach focuses on overcoming this HLS limitation by identifying an optimal mapping to both BRAMs and URAMs, and subsequently guiding the HLS tool according to this mapping (through $impl$=$\{BRAM/URAM\}$ in $pragma$ $bind\_storage$).

The inputs of our model include a MaxEVA solution ($X, Y, Z, M, K, N$ parameters), and PL-specific parameters, \emph{i.e.}, BRAM, URAM configurations supported in Versal devices \cite{Versal_acap_memory_resources_manual} and the available on-chip memory resources.
The model produces as outputs the optimal $U, V, W$ parameters, as well as the $A$, $B$ and $C$ buffer mapping to BRAMs/URAMs.

\begin{figure}[t]
\vspace{-0.8cm}
\centering
\subfloat[1K$\times$36 blocks]
{\includegraphics[width=0.52\linewidth]{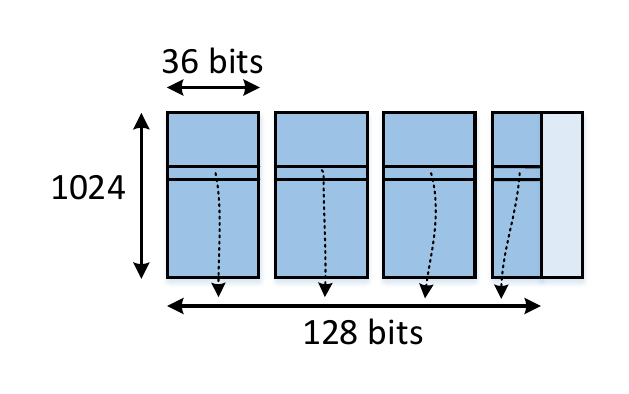}
\label{fig:1K_36}}
\subfloat[2 independent 1K$\times$18 blocks]{\includegraphics[width=0.48\linewidth]{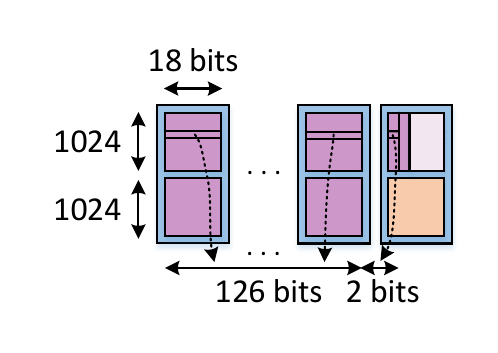}
\label{fig:2K_18}} 
\vspace{-0.1cm}
\caption{BRAM configurations example and proposed modeling.} 
\label{fig:BRAM_config_examples}
\vspace{-0.50cm}
\end{figure}

Besides the optimal parameter finding and mapping to memory resources, our model also estimates the BRAM/URAM utilization with 100\% accuracy in all cases (Sec. \ref{subsubsec:PL_model_estimation}).
First, we model the BRAM/URAM utilization based on the \textit{depth} of each partitioned buffer (all buffers have 128-bits width and are highly partitioned; Section \ref{subsubsec:PL_Implementation}).
Fig. \ref{fig:1K_36} shows an example of our modeling when $512 < depth \le 1K$.
In this case, BRAMs are configured as 1K$\times$36 and 4 36K BRAMs are needed to construct the 128-bitwidth buffer.
Observe that a portion of the rightmost BRAM in Fig. \ref{fig:1K_36} is not utilized in this situation.
To this end, we define BRAM/URAM \textit{efficiency} as the fraction of the \textit{logical} buffer size to the total size of memory blocks used.
For instance, when assuming that all 1K entries are used in Fig. \ref{fig:1K_36}, the BRAM \textit{efficiency} is determined by the utilized width, \emph{i.e.}, 128/(36$\cdot$4) = 88.89\%.

In Fig. \ref{fig:2K_18}, we show another example when $1K < depth \le 2K$.
In this case, BRAMs are configured as 2 \textit{independent} 1K$\times$18 blocks.
With 7 2K$\times$18 blocks, 126-bits can be mapped.
The remaining 2-bits can be efficiently mapped by packing 2K$\times$2-bits on a single 1K$\times$18 block.
This is possible since 2K$\times$2-bits can be \textit{logically} viewed as 1K$\times$4-bits, where additional multiplexing logic is needed to determine the corresponding 2-bits, based on the address.
Thus, 7.5 36K BRAMs (or 15 18K blocks) are 
required in total. 
The remaining 18K BRAM (outlined in orange in Fig. \ref{fig:2K_18}), can be used for other purposes, since it is physically independent.

Similarly, we also model the cases when  $depth \le 512$ (2 36K BRAMs), and $2K < depth \le 4K$ (15 36K BRAMs) \cite{Versal_acap_memory_resources_manual}.
In addition, since URAMs support one configuration (4K$\times$72), 2 URAMs are required for $depth \le 4K$.
In eq. \ref{eq:BRAMs_per_depth_function} and \ref{eq:URAMs_per_depth_function}, we summarize the number of BRAMs ($f_{B}$) and URAMs ($f_{U}$), respectively, as functions of the buffer depth.

\vspace{-3mm}
\small
\begin{align}
f_{B}(depth) & =
\begin{cases}
    2, & \text{if $depth \le 512$}\\
    4, & \text{if $ 512 < depth \le 1K$}\\
    7.5, & \text{if $ 1K < depth \le 2K$}\\
    15, & \text{if $ 2K < depth \le 4K$}\\   
\end{cases}
\label{eq:BRAMs_per_depth_function}
\\[3pt]
f_{U}(depth) & = 2, \quad  \text{if $depth \le 4K$}
\label{eq:URAMs_per_depth_function}
\end{align}
\normalsize
\vspace{-5mm}

Afterwards, we establish a constraint that limits each buffer's depth to 4K (eq. \ref{eq:depth_constraints}), since all buffers are highly partitioned with a relatively small depth.
This constraint allows a very high BRAM, URAM \textit{efficiency}, as shown in Sec. \ref{subsubsec:GEMM_performance_versal}. 
\begin{align}
\{A_{depth}, \ B_{depth}, \ C_{depth}\} \le 4K
\label{eq:depth_constraints}
\end{align}
Finally, we impose constraints to ensure that the total number of utilized BRAMs and URAMs does not exceed the device's available resources ($B_{36K}$ and $U_{288K}$).
Eq. \ref{eq:BRAMs_constraints_ex} and \ref{eq:URAMs_constraints_ex} show an example of the constraints for BRAMs and URAMs, respectively. 
In this example, the buffers $A$, $B$ have been mapped to BRAMs, while $C$ is mapped to URAMs.
\begin{gather}
A_{part} \cdot f_{B}(A_{depth}) +  B_{part} \cdot f_{B}(B_{depth}) \le B_{36K} \label{eq:BRAMs_constraints_ex}\\
C_{part} \cdot f_{U}(C_{depth}) \le U_{288K}
\label{eq:URAMs_constraints_ex}
\end{gather}
\noindent Constraints similar to  \ref{eq:BRAMs_constraints_ex}, \ref{eq:URAMs_constraints_ex} are applied for \textit{all} permutations of mapping buffers $A$, $B$ and $C$ to BRAMs and URAMs.

The solution of $U, V, W$ and buffer mapping to BRAMs, URAMs can be formulated as an integer programming (IP) optimization problem with the aforementioned constraints.
We solve the IP exhaustively by setting the maximization of on-chip data reuse as the objective.
The data reuse of all buffers is encapsulated in the product of $U$$\cdot$$V$$\cdot$$W$.
Notice from Fig. \ref{fig:GEMM_tiling_Versal} that buffers $A$ and $B$ are reused $W$ and $U$ times in GEMM, respectively, while $C$ is reused and updated (during accumulation) $V$ times.
In addition, note that maximizing data reuse also leads to maximization of the PL buffer sizes, under the resource constraints.
We report, implement and explore the trade-offs of multiple top-ranked solutions in Sec. \ref{subsec:Versal_results}.

\subsection{GEMM Implementation on Stratix 10 NX}
\label{subsec:GEMM_Architecture_Stratix}

We implement a configurable MatMul accelerator consisting of a control logic and a 2D TB layout on the Stratix 10 NX architecture. 
The accelerator's TBs operate out of local, on-chip memory consisting of $A$, $B$ and $C$ buffers (Fig.~\ref{fig_tensor_block_diagram}) of size $M^\prime$$\cdot$$K^\prime$, $K^\prime$$\cdot$$N^\prime$ and $M^\prime$$\cdot$$N^\prime$, respectively.
The 
$M^\prime$$\times$$K^\prime$$\times$$N^\prime$ MatMul size is defined as the \textit{native} buffer size, similar to Versal. 
We conduct a DSE to optimize for accelerator throughput and employ analytical modeling to optimize for on-chip data reuse.
Below we delineate the TB layout, the dataflow, the memory architecture, optimization strategies and the automatic code generation tool we developed.

\subsubsection{TB Layout}
\label{subsubsec:Stratix_TB_layout}
We utilize groups of TB \textit{arrays} of configurable length that run in parallel and operate in the cascade loading mode.
An \textit{array's} first TB (colored white in Fig.~\ref{fig_tb_layout}) serves as its point of entry for $A$ data (Fig.~\ref{fig_tb_layout}), performing no computation. 
Subsequent TBs in the \textit{array} receive their $A$ data (and store them in their registers) via their $casc\_data\_in$ port and their $B$ data through their $data\_in$ port (Fig.~\ref{fig_tensor_block_diagram}). The latency of loading $A$ blocks is three cycles per TB, but can be hidden when overlapped with dot-product operations. While $bank$ $0$ (Fig.~\ref{fig_tensor_block_diagram}) is being multiplied with a set of $B$ blocks, $bank$ $1$ can be loaded and vice-versa. An \textit{array's} accumulated outputs are exposed through the $data\_out$ port of its final TB.


\noindent We define four architecture parameters for the TB layout:
\paragraph{Parameter \arraylength} Length of a TB \textit{array}, equal to four in Fig.~\ref{fig_tb_layout}. 
This length can be less than or equal to 36.
\paragraph{Parameter \Kexpansion} Set of \textit{arrays} that work in parallel across the $K^\prime$ dimension, which we refer to as a \textit{reduction group} (\Kexpansion[] equal to two in Fig.~\ref{fig_tb_layout}). The $data\_out$ outputs of all the \textit{arrays} in a \textit{reduction group} are fed into its corresponding adder tree, the output of which is accumulated at the $C$ buffer (Fig.~\ref{fig_tb_layout}). These adders are implemented in soft logic. 
\paragraph{Parameter \Nexpansion} Set of \textit{reduction groups} that contain the same $A$ blocks, but get multiplied with different $B$ blocks (equal to two in Fig.~\ref{fig_tb_layout}). 
\Nexpansion[] allows exploiting parallelism across $N^\prime$. We refer to this set of \textit{reduction groups} as an \Nexpansion[] \textit{block}.
\paragraph{Parameter \Mexpansion} Number of \Nexpansion[] \textit{blocks} that allow parallelism across $M^\prime$, equal to three in Fig. \ref{fig_tb_layout}. Each \Nexpansion[] \textit{block} uses different $A$ blocks, but the same $B$ blocks.

\subsubsection{Dataflow}
\label{subsubsec:Stratix_TB_dataflow}
\begin{figure}[!t]
\centering
\vspace{-0.20cm}
\hspace{-0.40cm}
\includegraphics[width=0.98\linewidth]{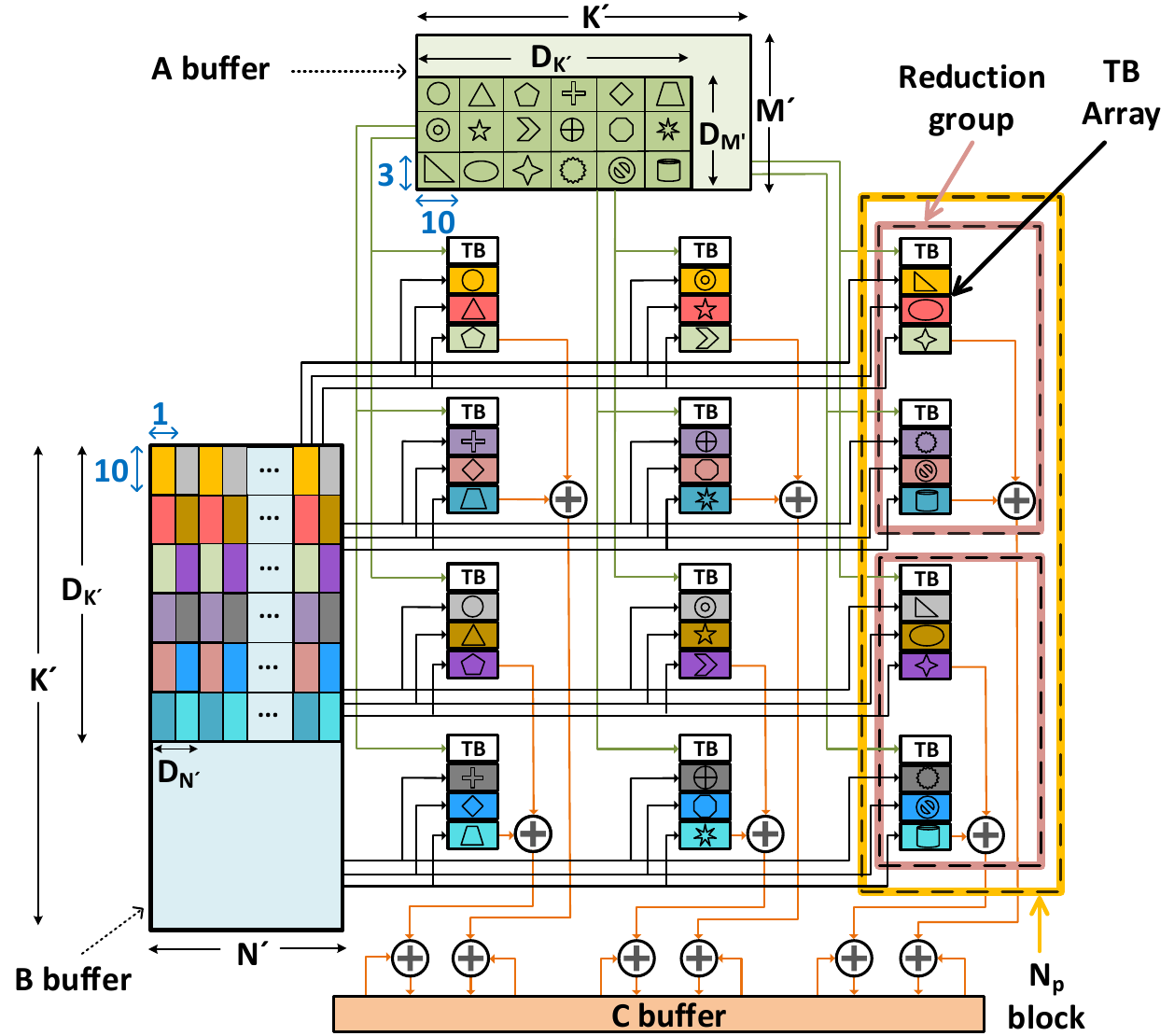}
\caption{2D TB layout and GEMM dataflow on Stratix 10 NX.}
\label{fig_tb_layout}
\vspace{-0.45cm}
\end{figure}

Each TB holds a 3$\times$10 block of $A$ in its registers, represented by a shape in Fig.~\ref{fig_tb_layout}, and it multiplies it with a 10$\times$1 block of $B$ in each clock cycle. This $B$ block belongs to a set of $N^\prime/\Nexpansion$ 10$\times$1 blocks, represented by a color in the $B$ Buffer (Fig.~\ref{fig_tb_layout}). The TB will sequentially multiply its $A$ block with all $B$ blocks in the colored set. Given enough $B$ blocks in this set (\emph{i.e.}, a large enough $N^\prime$), the TB's register loading latency can be fully hidden.
Different TBs of the same \textit{array} hold separate $A$ blocks (shapes) and process different sets of $B$ blocks (colors). Partial dot-products are propagated along an \textit{array} through cascade connections and added to the next dot-products on-the-fly. This process takes two clock cycles per TB (Sec.~\ref{subsec:Stratix_10_Nx_architecture}). Therefore, a TB starts processing data two cycles after its previous TB. Finally, an \textit{array's} accumulated dot-products are exposed at its last TB.

Different \textit{arrays} in the same \textit{reduction group} hold separate blocks of $A$ (shapes) and process separate sets of $B$ blocks (colors) in parallel, each generating a new output every clock cycle. 
\textit{Reduction groups} of the same \Nexpansion[] \textit{block} hold the same $A$ blocks (shapes), but get multiplied with different sets of $B$ blocks (colors) in parallel, allowing parallelism across $N^\prime$. Finally, different \Nexpansion[] \textit{blocks} contain different sets of $A$ blocks (shapes), but they all get multiplied with the same $B$ blocks (colors) in parallel, allowing parallelism across $M^\prime$.

The total number of utilized TBs is 
$\arraylength \cdot \Kexpansion \cdot \Nexpansion \cdot \Mexpansion$.
We define the \textit{compute} GEMM size as $(\Mexpansion\cdot3)$$\times$$([\arraylength-1]\cdot\Kexpansion\cdot10)$$\times$$(\Nexpansion)$. 
This is the size of the blocks the accelerator processes at the lowest memory hierarchy level. 
We refer to this GEMM size as $D_{M^\prime}$$\times$$D_{K^\prime}$$\times$$D_{N^\prime}$.
The multiplication by 3 and 10 accounts for the size of the $A$, $B$ blocks. We subtract by 1 for the wasted TB of each \textit{array} and we multiply by \Kexpansion[] for all \textit{arrays} in a \textit{reduction group}. 
Due to tiling over the \textit{compute} GEMM size, $M^\prime$, $K^\prime$ and $N^\prime$ need to be multiples of $D_{M^\prime}$, $D_{K^\prime}$ and $D_{N^\prime}$, respectively. 
Moreover, $N^\prime$ must be sufficiently large to allow hiding the TB register loading latency.

\subsubsection{Architectural Considerations}
\label{subsubsec:Stratix10_Architectural_considerations}
Below we elaborate on architectural considerations related to 
the
four architecture parameters, along with the constraints they introduce.

\paragraph{Parameter \arraylength[]} As mentioned, the TB \textit{chain} granularity is equal to 36. 
Hence, as multiple \textit{arrays} may fit in a chain, the \textit{array} length \arraylength[] must be a factor 
of 36 to avoid fragmentation during placement of a design with a large number
of TBs. 
Sound values for \arraylength[] are 36, 18, 12 and 9. Low \arraylength[] leads to more wasted TBs (first TB in every \textit{array}).
Meanwhile, this increases flexibility in executing different matrix sizes, as they do not require a high 
reduction dimension ($K^\prime$). Large \arraylength[] increases latency for loading $A$, but also results in higher peak throughput.  
A \arraylength[] of 36 has a loading latency of 36$\cdot$3=108 cycles. However, because each $A$ block is multiplied with a set of $B$ blocks, if this set contains at least 108 blocks, depending on $N^\prime$, 
this latency can be hidden. 
    
\paragraph{Parameter \Kexpansion} A TB \textit{array} processes $(\arraylength$$-$$1)$$\cdot$$10$ elements across the reduction dimension ($K^\prime$) each cycle. \Kexpansion[] increases the parallelism across $K^\prime$ (\emph{i.e.}, $D_{K^\prime}$), while keeping the signal fan-out low. 
However, if $D_{K^\prime}$ is large, but $K^\prime$ is small, some TBs in a \textit{reduction group} will be under-utilized.
    
\paragraph{Parameter \Nexpansion} 
In order to utilize more TBs, parallelism across $N^\prime$ and $M^\prime$ is also essential. \Nexpansion[] requires broadcasting $A$ blocks across all \textit{reduction groups} in an \Nexpansion[] \textit{block}, while \Mexpansion[] requires broadcasting $B$ blocks across all \Nexpansion[] \textit{blocks} (Fig.~\ref{fig_tb_layout}). However, each \textit{array} has only one input for $A$ data (at its initial TB), but a considerably larger amount of inputs for $B$ data (equal to $\arraylength$$-$$1$). As a result, \Mexpansion[] increases the fanout of a lot more signals than \Nexpansion[] does, which increases FPGA routing congestion.
Finally, a higher \Nexpansion[] leads to a larger minimum required value of $N^\prime$ to hide the TB loading latency.
    
\paragraph{Parameter \Mexpansion} 
\Mexpansion[] enables parallelism across $M^\prime$, but also increases the fan-out of $B$ blocks, which, as mentioned, can lead to significant routing congestion in the FPGA, adversely affecting attainable clock frequency and performance.

\subsubsection{Memory Architecture}
\label{subsubsec:Stratix_Memory}
Evidently,
when a large number of TBs are used, high BRAM bandwidth is needed to feed them blocks of $A$ and $B$. 
We employ a memory architecture that is designed to (i) enable high compute throughput and (ii) to maximize 
data reuse, which will amortize 
off-chip BW requirements. 
In a similar manner to Versal, we maintain separate buffers for $A$, $B$ and $C$ and we utilize double-buffering. 
The Load/Store units perform read-only operations on $C$ buffers and write-only operations on $A$ and $B$ buffers. 
The compute logic performs read and write operations on $C$ and read-only operations on $A$ and $B$. 
M20K blocks configured in \textit{simple} dual-port mode are used to implement all buffers. For $A$ and $B$, the compute and Load/Store units have their own port for reading and writing, respectively. For double-buffering, we double the depth of $A$ and $B$, and split $C$ to two equal-sized buffers.
A more detailed description of the architecture of buffers $A$, $B$ and $C$ is provided below.


\paragraph{Buffers $A$, $B$} As illustrated in Fig.~\ref{fig_tb_layout}, a separate $A$ block must be fed to each \textit{array} of a \textit{reduction group} and across different \Nexpansion[] \textit{blocks}. Additionally, each TB in an \textit{array} (except for $TB_0$), receives separate $B$ blocks every cycle. Similarly, all \textit{arrays} in all \textit{reduction groups} of an \Nexpansion[] \textit{block}  require different blocks of $B$ each clock cycle. Also, the bitwidth of the $A$ and $B$ ports is 80-bits. Therefore, the partition factors $A_{part}$, $B_{part}$ of $A$, $B$ and the depth $A_{depth}$, $B_{depth}$ of each smaller partitioned buffer of $A$, $B$ are:
\begin{gather} 
B_{part}=(\arraylength-1) \cdot \Kexpansion \cdot \Nexpansion \label{eq:NX_Bpart}\\
B_{depth} = 2\cdot K^\prime\cdot N^\prime/(B_{part}\cdot10)
\label{eq:NX_Bdepth}\\
A_{part}= \Mexpansion \cdot \Kexpansion, \ \ A_{depth} = 2\cdot M^\prime\cdot K^\prime/(A_{part}\cdot10)
\label{eq:NX_Apart_Adepth}
\end{gather}
We multiply by 2 for double-buffering and we also multiply by 10, because 80-bits is 10 bytes. 
The 80-bit width of these buffers is implemented using M20Ks operating in the 512$\times$40, 1024$\times$20 or 2048$\times$10 configurations~\cite{Intel_M20K_guide}. We model the number of M20Ks required to implement each buffer, $f_{M_{80}}$, as a function of the depth. For the $A$ buffer, since $A_{depth}$ is sufficiently large, all M20K configurations lead to equivalent, accurate solutions in our model. However, $B_{depth}$ was always less than 1024. If $depth \leq 512$, two M20Ks in 512$\times$40 mode can be used. If $512 < depth \leq 1024$, both 512$\times$40 and 1024$\times$20 configurations lead to a usage of four M20Ks. As a result, in all our designs $f_{M_{80}}$ is estimated as:

\small
\begin{equation} 
f_{M_{80}} (depth) = 2 \cdot \lceil depth/512 \rceil 
\label{eq:NX_fM40}
\end{equation}
\normalsize

\noindent The symbol $\lceil \ \rceil$ denotes rounding up to the next integer.

\paragraph{Buffer $C$} A TB has three $data\_out$ ports, so \textit{reduction groups} generate three distinct $C$ values each.
The partition factor $C_{part}$ and the depth $C_{depth}$ of each smaller buffer is:
\begin{equation}
    C_{part}= \Mexpansion \cdot \Nexpansion \cdot 3 \cdot 2, \ \ C_{depth} = M^\prime\cdot N^\prime\cdot2/C_{part}
    \label{eq:NX_Cpart_Cdepth}
\end{equation}
\noindent We multiply by 2 for double-buffering. 
The valid M20K configurations for the 32-bitwidth of the $C$ buffers are 2048$\times$8, 1024$\times$16 or 512$\times$32.
The number of M20Ks, $f_{M_{32}}$, required for $C$ buffers is also modeled as a function of the depth. Since the depth of the $C$ buffers is sufficiently large, all above M20K configurations led to equivalent, accurate solutions in our model, estimated by:

\small
\begin{equation} 
f_{M_{32}} (depth) = \lceil depth/512 \rceil 
\label{eq:NX_fM32}\\
\end{equation}
\normalsize

\subsubsection{Optimization Strategies}
\label{subsubsec:Stratix_Optimization}

Similar to Versal, we formulate the selection of $M^\prime$, $K^\prime$, $N^\prime$ for a given TB configuration as an IP problem and solve it exhaustively to optimize for data reuse, by maximizing the product $M^\prime$$\cdot$$K^\prime$$\cdot$$N^\prime$.
We impose two constraints on the IP. First, the utilized M20Ks must not exceed the device's available M20K blocks ($B_{M20K}$). Second, $N^\prime$ is set to be sufficiently large to hide TB register loading latency:
\begin{multline} 
    A_{part} \cdot f_{M_{80}}(A_{depth}) + B_{part} \cdot f_{M_{80}}(B_{depth})\\
    + C_{part} \cdot f_{M_{32}}(C_{depth}) \leq B_{M20K}
\label{eq:NX_M20K_constraint}
\end{multline}
\begin{equation}
    N^\prime \geq \arraylength \cdot 3 \cdot \Nexpansion[]
    \label{eq:Nprime_constraint} 
\end{equation}
Since throughput is directly related to operating frequency, we implement optimizations to shorten critical paths. These include replication of control logic and insertion of a configurable number of pipeline stages along the data and address datapaths.
Additionally, we conduct an extensive exploration on various TB architecture parameters, to find configurations that maximize frequency, and thus, throughput (Sec.~\ref{subsec:Stratix_results}).

In an effort to reduce signal fan-out, we also tried inserting registers between \Nexpansion[]{} \textit{blocks} in a systolic fashion in order to propagate $B$ blocks, thus minimizing broadcasting. However, our experiments showed no frequency improvements, while greatly increasing ALM usage (up to $\sim$40\%, as opposed to a maximum of $\sim$19\% with the proposed design, see Table~\ref{tab_NX_results}). Such a large number of soft-logic pipeline registers is undesirable, as it can reduce energy efficiency and also render logic resources unavailable, as mentioned in~\cite{langhammer2021stratix}. Thus, the systolic distribution approach was not considered in our final designs.



\subsubsection{Automatic RTL Code Generation}
\label{subsubsec:Stratix_Design_Generation}
We develop a Python-based tool that automatically generates the aforementioned architecture's RTL code, including the control logic, TB layout and memory. The input configurations of the tool are the four architecture parameters, the 
$M^\prime$, $K^\prime$, $N^\prime$  dimensions, 
and the number of pipeline stages for both address and data.

%% file: Results/Results.tex
\subsection{Versal DSE Evaluation}
\label{subsec:Versal_results}

For the AIE, we obtain the two most efficient solutions from MaxEVA.
As found in \cite{Taka2023MaxEVA}, solution $P1$ 13$\times$4$\times$6 ($X$$\times$$Y$$\times$$Z$) shows the highest throughput, while the $P2$ 10$\times$3$\times$10 presents the highest energy efficiency.
Both use a 32$\times$128$\times$32 ($M$$\times$$K$$\times$$N$) single AIE MatMul kernel with 95\% throughput efficiency.
For the PL, we utilize our model (Sec. \ref{subsubsec:PL_Optimization}), which takes as input a MaxEVA solution ($X, Y, Z, M, K, N$ parameters) and produces the optimal values of $U, V, W$.
We perform DSE on the 5 top-ranked $U$$\times$$V$$\times$$W$ solutions, for each of the two MaxEVA solutions (10 designs in total).

We use the AMD/Xilinx Vitis 2022.1 version to implement and compile our designs.
The PL part is designed using Vitis HLS, while AIE-PL linking is achieved via the V++ compiler.
Throughout all experiments, the AIE frequency is set to its maximum value of 1.25 GHz, while the PL frequency ranges from 275--300 MHz, depending on the PL configuration.
To calculate the throughput of our designs, we use hardware emulation in Vitis, while
power is estimated through the post-implementation Vivado Power Analysis Tool \cite{Vivado_power_analysis}.

\subsubsection{Model Estimation}
\label{subsubsec:PL_model_estimation}

In Table \ref{tb:HLS_results}, we present 4 top-ranked solutions of our PL optimization procedure (parameters $U$$\times$$V$$\times$$W$ and buffer mapping to BRAM/URAM resources).
First, we observe that our model estimates the BRAM/URAM utilization with 100\% accuracy in all cases.
For example, for solution 4$\times$2$\times$4 ($P1$), the model suggests (``Model Est." column) that buffers $\{A, B, C\}$ should be mapped to $\{\mathcal{B, U, U}\}$, where $\mathcal{B}$, $\mathcal{U}$ denote BRAM, URAM, respectively.
When we guide the HLS tool to use this mapping, 
both our model and HLS synthesis report identical BRAM/URAM utilization, \emph{i.e.}, 780 (81\%) / 408 (88\%).
Letting the HLS tool instead to automatically map buffers (``HLS AUTO" column), results in a severe over-utilization of URAMs (616 or 133\%), without any BRAM usage.
Vivado PnR attempts to implement this HLS AUTO solution by mapping the surplus URAMs to BRAMs.
However, it generates an error reporting a 119.4\% BRAM and 99.8\% URAM utilization.
A similar result is observed for another solution, \emph{i.e.}, 4$\times$2$\times$4 ($P2$).
However, for the other two solutions shown in Table \ref{tb:HLS_results}, HLS AUTO is able to successfully find an efficient mapping.
In these cases, both our model and HLS AUTO produce exactly the same BRAM/URAM utilization, \emph{e.g.}, 416 (43\%) / 408 (88\%), for 2$\times$2$\times$8 ($P1$).
We note that HLS AUTO fails to implement 5 of the 10 top solutions (Table \ref{tb:Versal_top10_designs}), justifying the necessity of our approach.


\subsubsection{GEMM Performance}
\label{subsubsec:GEMM_performance_versal}
In Table \ref{tb:Versal_top10_designs}, we show various metrics for our 10 top solutions.
In all solutions, throughput is calculated on their \textit{native} buffer sizes 
(Sec. \ref{subsubsec:GEMM_Tiling_Versal})
Overall, we observe that all designs exhibit high throughput ranging from 75.4--77.01 TOPs.
However, to maintain such high throughput, several designs require higher DDR BW compared to the VC1902's BW (102.4 GB/s).
We note that we calculate DDR BW
as the worst-case of concurrent loads for buffers $A$, $B$ and stores for $C$ (all as 8-bits due to quantization in DL).
With more sophisticated data reuse techniques as in \cite{XVDPU_AIE_FPL22}, this requirement can be amortized. 
In this work, we consider the worst-case BW scenario for the sake of generality.
Therefore, we only examine designs that stay within the BW of the VC1902 device (highlighted in bold in Table \ref{tb:Versal_top10_designs}).

From these designs, 2$\times$2$\times$8 ($P1$) shows both the highest throughput, \emph{i.e.}, 76.93 TOPs, and the best energy efficiency, \emph{i.e.}, 0.938 TOPs/W.
For all valid solutions, throughput ranges from 75.40--76.93 TOPs, which is 58.9--60.1\% of the \textit{theoretical peak} throughput of VC1902, and the same as the state-of-the-art MaxEVA~\cite{Taka2023MaxEVA}.
Additionally, energy efficiency ranges from 0.911--0.938 TOPS/W.
In all cases, we notice a very high resource utilization, up to 94\% BRAMs, 88\% URAMs, and 100\% AIE cores, with a small LUT usage of up to 11\%.
Moreover, we observe high RAM \textit{efficiency} of 75.7--90.2\%, as a direct result of our modeling and optimization methodology.
Finally, although not shown in Table \ref{tb:Versal_top10_designs} for brevity, the swapping of the first and last GEMM dimensions results in equivalent solutions (symmetrical design, see Sec. \ref{subsubsec:PL_Implementation}).
For instance, solution 2$\times$2$\times$8 ($P1$) can be swapped to 8$\times$2$\times$2 ($P1$), resulting in a \textit{native} buffer size of 1536$\times$1024$\times$832 (\textit{compute} GEMM size becomes 192$\times$512$\times$416 in this case).

\begin{table}[t]
 \centering
\caption{Optimization model estimation for various solutions and comparison with HLS AUTO mapping.}
\setlength\tabcolsep{2.6pt}
\renewcommand{\arraystretch}{1.10}
\begin{tabular}{c|ccc|cc}
\Xhline{2.5\arrayrulewidth}

\textbf{U$\times$V$\times$W}  &  
\multicolumn{3}{c|}{\textbf{Model Estimation}} & \multicolumn{2}{c}{\textbf{HLS AUTO}}\\

\cline{2-4} \cline{5-6}
\textbf{(MaxEVA P.)} & \textbf{\textbf{\{A, B, C\}}} & \textbf{BRAMs} & \textbf{URAMs} & \textbf{BRAMs} & \textbf{URAMs}\\
\hline
\hline
4$\times$2$\times$4 (P1) & $\{\mathcal{B, U, U}\}$ & 780 (81\%) & 408 (88\%) & \textbf{0 (0\%)} & \textbf{616 (133\%)}\\

4$\times$2$\times$4 (P2) & $\{\mathcal{B, B, U}\}$ & 900 (93\%) & 400 (86\%) & \textbf{0 (0\%)} & \textbf{640 (138\%)}\\

2$\times$2$\times$8 (P1) & $\{\mathcal{B, U, U}\}$ & 416 (43\%) & 408 (88\%) & 416 (43\%) & 408 (88\%)\\

2$\times$8$\times$2 (P2) & $\{\mathcal{U, U, B\}}$ & 800 (83\%) & 240 (52\%) & 800 (83\%) & 240 (52\%)\\











\Xhline{2.5\arrayrulewidth}

\end{tabular}

\label{tb:HLS_results}
\vspace{-0.30cm}
\end{table}

\subsection{Stratix DSE Evaluation}
\label{subsec:Stratix_results}


\begin{table*}[t]
 \centering
\caption{Evaluation of 10 top-ranked GEMM designs on Versal VC1902. AIE operates at 1.25 GHz.}
\vspace{-0.10cm}
\setlength\tabcolsep{3.5pt}
\renewcommand{\arraystretch}{1.12}
\resizebox{1.00\textwidth}{!}{
\vspace{-0.10cm}
\begin{tabular}{c|c|c|c|c|c|c|c|c|c|c|c|c}
\Xhline{2.5\arrayrulewidth}

\textbf{U$\times$V$\times$W} & \textbf{Compute} & \textbf{Native} & \multirow{2}{*}{\textbf{LUTs}}  &  \multirow{2}{*}{\textbf{BRAMs}}  & \multirow{2}{*}{\textbf{URAMs}}  & \textbf{AIE} & \textbf{PL Fq.} & \textbf{Thrpt.} & \textbf{Power} & \textbf{En. Eff.} & \textbf{RAM} & \textbf{BW}\\

\textbf{(MaxEVA P.)} & \textbf{GEMM size} & \textbf{Buffer size}  &    &   &   & \textbf{cores} & \textbf{(MHz)} & \textbf{(TOPs)} & \textbf{(W)} & \textbf{(TOPs/W)} & \textbf{Eff.} & \textbf{(GB/s)}\\

\hline
\hline

2$\times$8$\times$2 (P1) & 416$\times$512$\times$192 & 832$\times$4096$\times$384 & 85K (9\%) & 630 (65\%) & 304 (66\%) & 390 (98\%) & 300 & 77.01 & 78.6 & 0.980 & 88.9\% & 145.2\\

\textbf{2$\times$2$\times$8 (P1)} & \textbf{416$\times$512$\times$192} & \textbf{832$\times$1024$\times$1536} & \textbf{91K (10\%)} & \textbf{422 (44\%)} & \textbf{408 (88\%)} & \textbf{390 (98\%)} & \textbf{290} & \textbf{76.93} & \textbf{82.0} & \textbf{0.938 }& \textbf{88.9\%} & \textbf{101.4}\\

\textbf{3$\times$2$\times$5 (P1)} & \textbf{416$\times$512$\times$192} & \textbf{1248$\times$1024$\times$960} & \textbf{94K (10\%)} &\textbf{ 792 (82\%)} & \textbf{408 (88\%)} & \textbf{390 (98\%)} & \textbf{278} & \textbf{76.72} & \textbf{82.7} & \textbf{0.932} & \textbf{75.7\%} & \textbf{100.7}\\

\textbf{4$\times$2$\times$4 (P1)} & \textbf{416$\times$512$\times$192} & \textbf{1664$\times$1024$\times$768} & \textbf{90K (10\%)} & \textbf{792 (82\%)} & \textbf{408 (88\%)} & \textbf{390 (98\%)} & \textbf{278} & \textbf{76.72} & \textbf{82.3 }& \textbf{0.928} & \textbf{81.6\%} & \textbf{101.9}\\

2$\times$4$\times$4 (P1) & 416$\times$512$\times$192 & 832$\times$2048$\times$768 &  97K (11\%) & 792 (82\%) & 408 (88\%) & 390 (98\%) & 278 & 76.72 & 82.8 & 0.927 & 62.6\% & 106.9\\

\hline

2$\times$8$\times$2 (P2) & 320$\times$384$\times$320 & 640$\times$3072$\times$640 &  92K (10\%) & 806 (83\%) & 240 (52\%) & 400 (100\%) & 300 & 76.08 & 78.3 & 0.971 & 88.9\% & 122.2\\

2$\times$7$\times$2 (P2) & 320$\times$384$\times$320 & 640$\times$2688$\times$640 & 92K  (10\%) & 806 (83\%) & 240 (52\%) & 400 (100\%) & 300 & 76.08 & 77.8 & 0.977 & 81.0\% & 123.9\\

2$\times$6$\times$2 (P2) & 320$\times$384$\times$320 & 640$\times$2304$\times$640 & 91K (10\%) & 806 (83\%) & 240 (52\%) & 400 (100\%) & 300 & 76.08 & 77.5 & 0.982 & 73.2\% & 126.1\\

\textbf{4$\times$2$\times$4 (P2)} & \textbf{320$\times$384$\times$320} & \textbf{1280$\times$768$\times$1280} & \textbf{100K (11\%)} & \textbf{912 (94\%)} & \textbf{400 (86\%)} & \textbf{400 (100\%)} & \textbf{275} & \textbf{75.40} & \textbf{82.8} & \textbf{0.911}  & \textbf{90.2\%} & \textbf{100.6}\\

4$\times$2$\times$3 (P2) & 320$\times$384$\times$320 & 1280$\times$768$\times$960 & 100K (11\%) & 912 (94\%) & 400 (86\%) & 400 (100\%) & 275 & 75.40 & 82.0 & 0.919 & 70.2\% & 109.7\\

\Xhline{2.5\arrayrulewidth}

\end{tabular}
}
\label{tb:Versal_top10_designs}
\end{table*}

\begin{table*}[h]
\centering
\setlength\tabcolsep{3.8pt}
\renewcommand{\arraystretch}{1.04}
\caption{Evaluation of 10 top-ranked GEMM designs on Stratix 10 NX.}
\vspace{-0.10cm}
\begin{tabular}{c|c|c|c|c|c|c|c|c|c|c|c}
\Xhline{2.5\arrayrulewidth}
\multirow{2}{*}{\textbf{TB config.}} & \textbf{Compute} & \textbf{Native} & \multirow{2}{*}{\textbf{ALMs}} & \multirow{2}{*}{\textbf{BRAMs}} & \multirow{2}{*}{\textbf{TBs}} & \textbf{Freq.} & \textbf{Thrpt.} & \textbf{Power} & \textbf{En. Eff.} & \textbf{RAM} & \textbf{BW}\\
 & \textbf{GEMM size} & \textbf{Buffer size} &  &  &  & \textbf{(MHz)} & \textbf{(TOPs)} & \textbf{(W)} & \textbf{(TOPs/W)} & \textbf{Eff.} & \textbf{(GB/s)}\\

\hline
\hline
18$\times$16$\times$4$\times$3         & 9$\times$2720$\times$4           & 639$\times$2720$\times$1008          & 124K (18\%)          & 6304 (92\%)          & 3456 (87\%)          & 349          & 68.00          & 51.1          & 1.331          & 88.0\%          & 92.6          \\
18$\times$8$\times$8$\times$3          & 9$\times$1360$\times$8           & 675$\times$2720$\times$928           & 123K (17\%)          & 6064 (89\%)          & 3456 (87\%)          & 345          & 67.21          & 50.2          & 1.340          & 87.7\%          & 91.6          \\
\textbf{9$\times$16$\times$5$\times$5} & \textbf{15$\times$1280$\times$5} & \textbf{900$\times$1280$\times$1000} & \textbf{127K (18\%)} & \textbf{5840 (85\%)} & \textbf{3600 (91\%)} & \textbf{350} & \textbf{66.94} & \textbf{52.5} & \textbf{1.275} & \textbf{81.2\%} & \textbf{90.2} \\
12$\times$8$\times$6$\times$6          & 18$\times$880$\times$6           & 1152$\times$1760$\times$756          & 100K (14\%)          & 6144 (90\%)          & 3456 (87\%)          & 338          & 64.00          & 48.6          & 1.317          & 86.7\%          & 82.2          \\
18$\times$16$\times$3$\times$4         & 12$\times$2720$\times$3          & 850$\times$2720$\times$750           & 108K (15\%)          & 6272 (92\%)          & 3456 (87\%)          & 327          & 63.71          & 47.3          & 1.347          & 85.9\%          & 85.4          \\
9$\times$16$\times$6$\times$4          & 12$\times$1280$\times$6          & 912$\times$2560$\times$756           & 131K (19\%)          & 6464 (94\%)          & 3456 (87\%)          & 342          & 62.88          & 50.7          & 1.241          & 85.1\%          & 82.3          \\
18$\times$8$\times$3$\times$8          & 24$\times$1360$\times$3          & 1600$\times$1360$\times$550          & 81K (12\%)           & 6064 (89\%)          & 3456 (87\%)          & 321          & 62.40          & 46.5          & 1.342          & 83.1\%          & 92.4          \\
\textbf{9$\times$8$\times$10$\times$5} & \textbf{15$\times$640$\times$10}  & \textbf{900$\times$1280$\times$1000} & \textbf{124K (18\%)} & \textbf{5840 (85\%)} & \textbf{3600 (91\%)} & \textbf{320} & \textbf{61.21} & \textbf{48.7} & \textbf{1.257} & \textbf{81.2\%} & \textbf{82.4} \\
18$\times$8$\times$5$\times$5          & 15$\times$1360$\times$5          & 1020$\times$2720$\times$630          & 101K (14\%)          & 6150 (90\%)          & 3600 (91\%)          & 301          & 61.08          & 45.4            & 1.346          & 90.0\%          & 83.5          \\
18$\times$4$\times$8$\times$6          & 18$\times$680$\times$8           & 1152$\times$1360$\times$832          & 91K (13\%)           & 6080 (89\%)          & 3456 (87\%)          & 312          & 60.69          & 46.2         & 1.315          & 84.3\%          & 79.3
\\
\Xhline{2.5\arrayrulewidth}
\end{tabular}
\label{tab_NX_results}
\vspace{-0.25cm}
\end{table*}

We conducted a DSE on 100 designs with different TB architecture parameters 
to optimize for performance and energy efficiency. The TB utilization of all explored designs is 85\%--91\%. The \textit{native} buffer sizes (Sec. \ref{subsec:GEMM_Architecture_Stratix}) were set based on our IP solver. 
The RTL (Verilog) code 
was automatically generated using our Python tool, and afterwards implemented on Intel Quartus 2021.1 (with a patch for Stratix 10 NX support from Intel). We use 
ModelSim~\cite{ModelSim_user_guide} to calculate throughput and the Quartus Power Analyzer to estimate power~\cite{Quartus_power_analysis}.

Table \ref{tab_NX_results} shows the evaluation of the top 10 designs.
Configurations are a combination of the 4 TB architecture parameters ($\arraylength\times\Kexpansion\times \Nexpansion\times\Mexpansion$). Designs are ranked based on their throughput when running GEMM on their \textit{native} buffer sizes.
Our solutions achieve high throughput, up to 68 TOPs, which is 47.6\% of the \textit{theoretical peak} of the NX 2100 device. 
We note that our throughput is directly related to our achieved frequencies, which are similar to prior work~\cite{hpipex_NX,boutros2020beyond,langhammer2021stratix}. The maximum achieved energy efficiency is 1.347 TOPS/W. All designs present a small ALM  (12-19\%), but very high BRAM (85--94\%) and TB (87--91\%) utilization.
Moreover, they present very high RAM \textit{efficiency} (81.2--90\%) and low BW requirements (79.3--92.6 GB/s).

Overall, we notice that the top designs have a \arraylength[] of 18, 12 or 9. Designs with a \arraylength[] of 36 achieve lower frequency and require a longer time for PnR, as they have less flexibility during placement.
Note that design 9$\times$16$\times$5$\times$5 has a lower throughput than 18$\times$16$\times$4$\times$3, despite having higher TB utilization and operating at a higher frequency.
This is attributed to a smaller \arraylength[], and therefore more wasted TBs operating in parallel load mode. Notice also how designs 18$\times$16$\times$4$\times$3 and 18$\times$8$\times$8$\times$3 perform better than designs 18$\times$16$\times$3$\times$4 and 18$\times$8$\times$3$\times$8, respectively. 
They achieve a higher frequency, due to a lower \Mexpansion[], which leads to lower overall signal fan-out and FPGA routing congestion. 


\subsection{Insights \& Discussion}
\label{subsec:Comparison}

\begin{figure}[t]
\vspace{-0.4cm}
\centering
\subfloat[]{\includegraphics[width=0.50\linewidth]{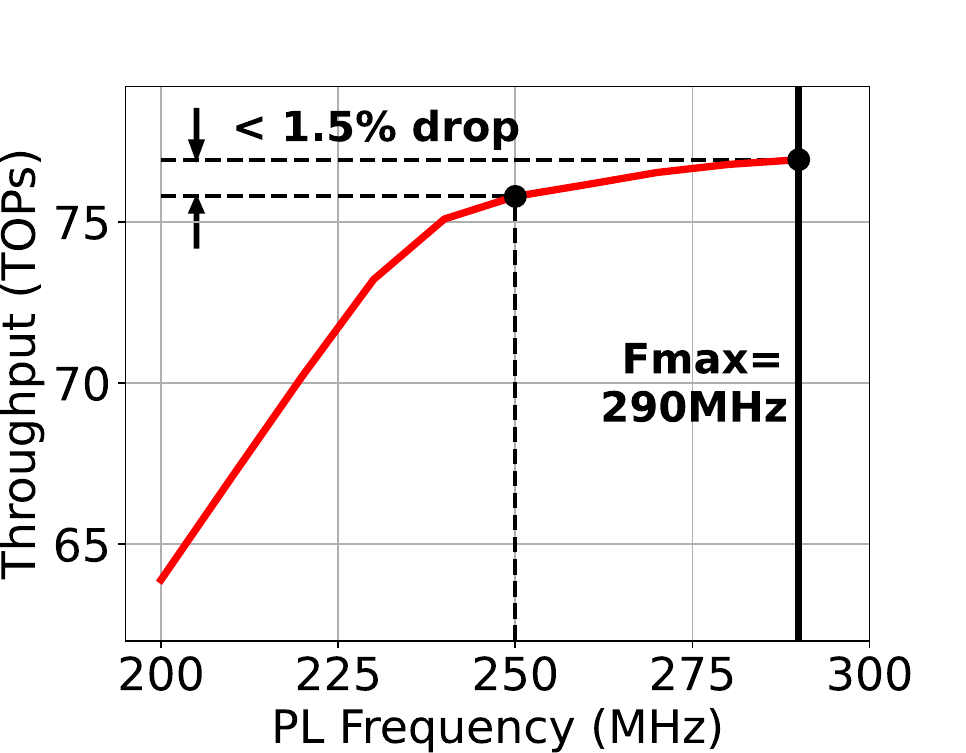}
\label{fig:throughput_PL_freq}}
\subfloat[]{\includegraphics[width=0.50\linewidth]{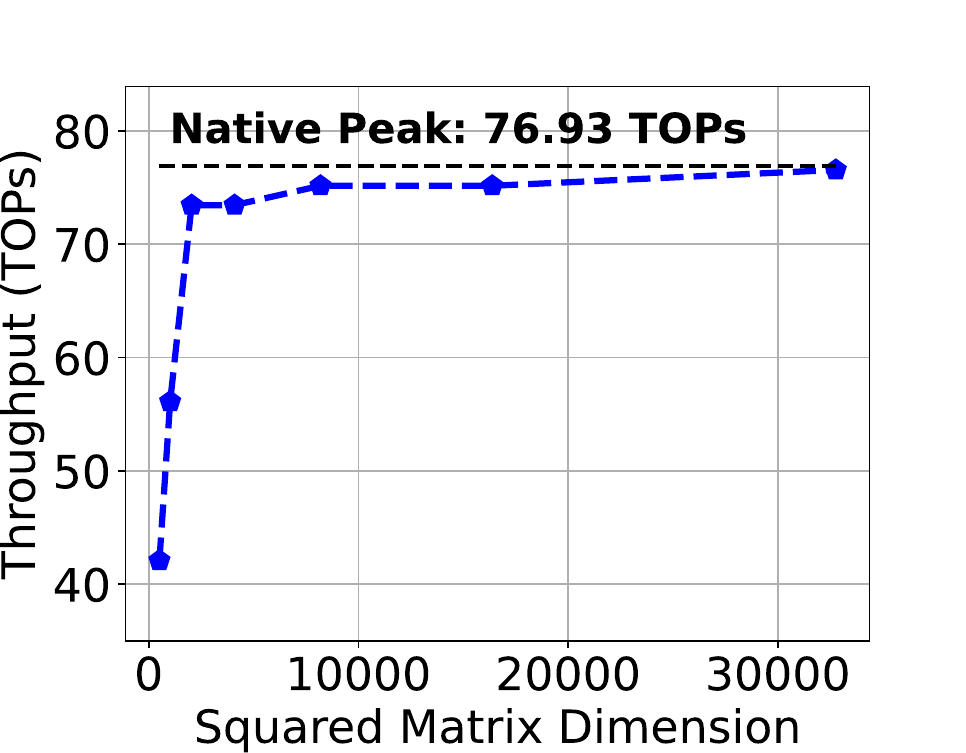}
\label{fig:throughput_mat_size_2x2x8_P1}}
\caption{Performance \emph{vs.} PL frequency (a), and square matrix dimensions (b) of 2$\times$2$\times$8 ($P1$) on Versal VC1902.}
\label{fig:throughput_per_matrix_size_PL_freq}
\vspace{-0.40cm}
\end{figure}

\noindent \textbf{\textit{1) FPGA Frequency:}} We examine the reliance of our solutions on the attainable FPGA frequency. While Versal's AIE array operates at a fixed frequency (1.25 GHz), the PL (FPGA) frequency varies from 275-300 MHz across all top solutions (Table \ref{tb:Versal_top10_designs}).
Despite the PL frequency variation, we observe a 
stable performance ($\sim$2\%) for Versal.
We further depict the impact of PL frequency on throughput in Fig. \ref{fig:throughput_PL_freq} for design 2$\times$2$\times$8 ($P1$).
We notice a negligible throughput decrease ($<$1.5\%)  for a wide frequency range (290 to 250 MHz).
This is ascribed to higher AIE computation time compared to communication with PL, illustrating the \textit{weak} dependence of performance on a \textit{wide} PL frequency range. 
However, for frequencies lower than 250 MHz performance degrades more severely ($\sim$16\% from 250 to 200 MHz).

Conversely, for Stratix, we observed a higher performance range ($\sim$12\%) across all solutions in Table \ref{tab_NX_results}.
This is attributed to the direct relationship of performance to frequency, illustrating the \textit{strong} dependence of NX on PL frequency.




\noindent \textbf{\textit{2) GEMM Scalability}}: Furthermore, we explore the scalability of our solutions when altering the matrix dimensions.
For Versal, in Fig. \ref{fig:throughput_mat_size_2x2x8_P1}, we present the throughput of the best overall design 2$\times$2$\times$8 ($P1$), for square matrices (powers-of-2) ranging from 512 to 32K.
Zero-padding is applied to align to the \textit{compute} GEMM size (Table \ref{tb:Versal_top10_designs}).
We notice that our design scales effectively, leading to almost its \textit{native} peak 
throughput for dimensions $\sim$2K and higher.
We note here that \textit{native} peak refers to the \textit{achieved} throughput of our designs when running GEMM on their \textit{native} buffer sizes (Tables \ref{tb:Versal_top10_designs}, \ref{tab_NX_results}).

\begin{figure}[t]
\vspace{-0.4cm}
\centering
\subfloat[]{\includegraphics[width=0.50\linewidth]{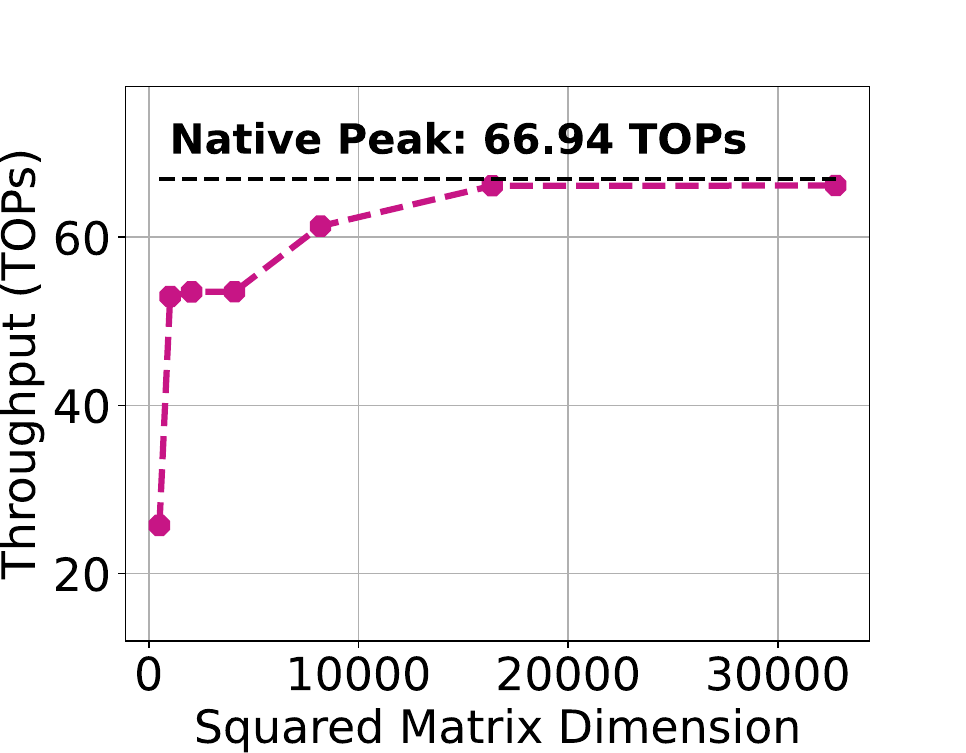}
\label{fig:high_K_design_NX}}
\subfloat[]{\includegraphics[width=0.50\linewidth]{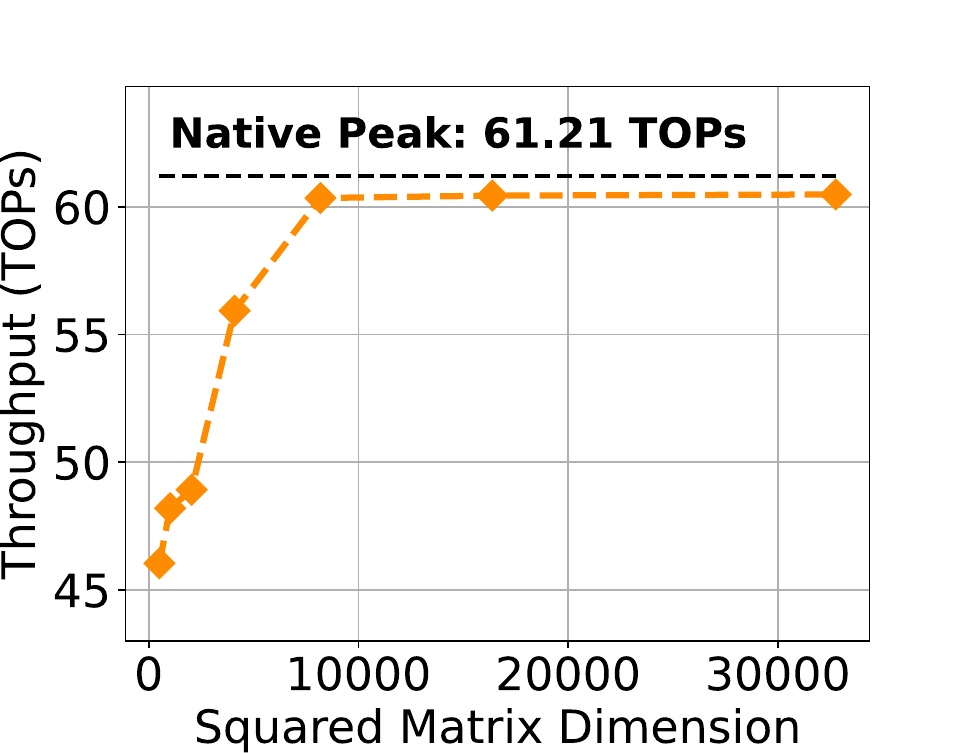}
\label{fig:low_K_design_NX}}
\caption{Performance \emph{vs.} square matrix dimensions of designs  9$\times$16$\times$5$\times$5 (a) and 9$\times$8$\times$10$\times$5 (b) on Stratix 10 NX.}
\label{fig:stratix10_NX_results}
\vspace{-0.40cm}
\end{figure}

Similarly, for Stratix, we show the scalability of the two bolded designs in Table~\ref{tab_NX_results}.
The first (Fig. \ref{fig:high_K_design_NX}) has a high $D_{K^\prime}$, and presents one of the highest throughput among all top solutions.
The second (Fig. \ref{fig:low_K_design_NX}) has the lowest $D_{K^\prime}$. 
As $D_{K^\prime}$ is much higher than $D_{M^\prime}$ and $D_{N^\prime}$ (Table~\ref{tab_NX_results}), designs with flexible (small) $D_{K^\prime}$ require less zero-padding along the reduction dimension ($K^\prime$) and scale better.
We observe that, while the design in Fig.~\ref{fig:high_K_design_NX} has higher \textit{native} peak throughput, the design in Fig.~\ref{fig:low_K_design_NX} scales better, due to lower $D_{K^\prime}$.
Overall, GEMM scalability on both devices directly depends on the \textit{compute} GEMM size. 
By properly adjusting this size, our methods can be exploited in straightforward fashion to target specific matrix dimensions \emph{e.g.}, long and narrow matrices. 
However, for the sake of generality, we demonstrate GEMM scalability on square matrices, without targeting specific sizes.





\noindent \textbf{\textit{3) Achieved Performance}}:
As mentioned, Versal achieves $\sim$60\% of its \textit{theoretical peak} throughput. 
This can be ascribed to multiple reasons.
First, the non-ideal (95\%) efficiency~\cite{Taka2023MaxEVA, charm2023fpga} of the AIE MatMul kernels.
Second, the necessity to introduce Add kernels on the AIE, which do not contribute to throughput, but occupy some of the cores (Sec. \ref{subsubsec:GEMM_Mapping_AIE}).
Third, the inevitable stalls caused by memory conflicts on the AIE array \cite{AI_Engine_programming_environment}.
In contrast, Stratix achieves $\sim$47\% of its \textit{theoretical peak} throughput. This is because the \textit{theoretical peak} assumes 100\% TB utilization with no wasted TBs in cascade mode, and a very high operating frequency of 600MHz~\cite{langhammer2021stratix}, which are infeasible to attain in practice.

Our results in Tables \ref{tb:Versal_top10_designs}, \ref{tab_NX_results} indicate that, on average, Versal attains 19.8\% higher throughput, while Stratix presents 41.6\% higher energy efficiency on GEMM workloads.
However, we note that the focus of this study strays from a competitive comparison among the two AI-optimized FPGA architectures.
Instead, we emphasize the distinct design
methods required for each device, due to their considerably different architecture styles (out-of-fabric \emph{vs.} in-fabric).
A broader set of full AI-workloads would be needed to explore complicated trade-offs between the two architectures, which we leave as future work.

\noindent \textbf{\textit{4) Programmability Trade-Offs}}: The two technologies require completely different programming methods.
In particular, Versal utilizes high-level C/C++ software for programming the AIE array. 
Efficient, high-level programming constructs can enable high performance of the AIE array's SIMD processors on vectorizable workloads (with 95\% AIE kernel efficiency).
For productivity purposes, the PL part is typically programmed in HLS. 
However, HLS inefficiencies can lead to performance degradation, which can be 
mitigated by exploiting sophisticated methods (Sec. \ref{subsubsec:PL_Optimization}).
Finally, AMD/Xilinx provides automation tools, such as Vitis V++, to reduce the complexity of integrating the entire system.


On the contrary, no high-level tools are available for programming Stratix 10 NX.
In particular, Intel HLS~\cite{Intel_HLS_guide} has no support for TBs, thus requiring exclusive coding in RTL for design, verification, and full-system integration.
This resulted in $\sim$40$\times$ higher number of lines of code in Stratix compared to Versal.
Nonetheless, our Python-based generator significantly reduced this complexity.
Moreover, because the performance of Stratix is directly analogous to frequency, unlike Versal, more programming effort is required to meet timing goals, and an extensive DSE is needed to identify high-throughput designs. In our case, this led to $\sim$5$\times$ higher total tool compilation time for Stratix compared to Versal.
While each individual design required 3--6 hours to compile on both devices, the extensive exploration on Stratix greatly increased the design space. Finally, based on our estimations, it might be 1.5--2$\times$ more productive to program Versal due to the 
increased programming
effort required by Stratix. However, we note that design productivity is highly dependent on designer skills.





\noindent \textbf{\textit{5) GEMM Optimization Insights}}:
Both devices exhibit design challenges due to their high complexity.
To address complexity, systematic methodologies, \emph{e.g.}, analytical modeling, are essential for optimization of GEMM-based applications.

Versal AIE is a novel architecture that introduces several new design challenges.
A comprehensive exploitation of AIE architectural attributes, \emph{e.g.}, local memory sharing and circuit switching, is crucial to optimize performance in GEMM \cite{Taka2023MaxEVA}.
Moreover, challenges, such as the reduced number of AIE-PL tiles and AIE routing congestion, require refined techniques to prevent performance degradation (Sec. \ref{subsubsec:GEMM_Mapping_AIE}).
Finally, the limited DRAM BW of Versal devices introduces additional challenges in maintaining the high throughput of the AIEs.


In contrast, Stratix 10 NX preserves a traditional FPGA architecture, while introducing TBs embedded in the PL.
The inherent complexity of TBs necessitates the
consideration of multiple operating modes to effectively map GEMM workloads (Sec.~\ref{subsec:GEMM_Architecture_Stratix}).
Moreover, as throughput is directly related to frequency, performance depends on both low-level RTL optimizations and the intricacies of the FPGA architecture. RTL techniques, \emph{e.g.}, replication of control logic and insertion of pipeline stages, are essential in shortening critical paths. 

%% file: Summary-Future_work/Summary_Future_Work.tex
In this work, we propose novel methodologies/frameworks to optimize GEMM-based applications targeting the two leading AI-optimized FPGA architectures.
We present a thorough evaluation of several aspects in GEMM design, by efficiently leveraging the unique and substantially different architectural attributes of the AMD/Xilinx Versal ACAP and Intel Stratix 10 NX devices.
Our experimental results show that our frameworks achieve up to 77 and 68 TOPs throughput on GEMM workloads for Versal and Stratix, respectively.
This study provides fundamental insights regarding the architectural characteristics, programmability trade-offs, challenges and limitations inherent in both Versal and Stratix AI-optimized FPGAs.